\documentclass{aa}  

\usepackage{graphicx}
\usepackage{txfonts}
\usepackage{lscape}
\usepackage{hyperref}
\begin{document} 

   \title{Intermediate-redshift slow rotators as the progenitors of the most massive galaxies at z $\sim$ 0}

   \author{C. Mu\~{n}oz L\'{o}pez
          \inst{1}\fnmsep\thanks{e-mail: \texttt{cmunoz@aip.de}}
          Davor Krajnovi\'{c}
          \inst{1}
          \and
          B. Epinat
          \inst{2,3}
          \and
          T. Contini
          \inst{4}
          \and
          I. Pessa
          \inst{1}
          \and
          J. Pharo
          \inst{1}
          \and
          A. de Araujo Carvalho
          \inst{5}
          \and
          A. R. Chougule
          \inst{6,7}
          .
          }

   \institute{Leibniz-Institut f\"ur Astrophysik Potsdam (AIP),
        An der Sternwarte 16, 14482 Potsdam, Germany
        \and
        Aix-Marseille Univ., CNRS, CNES, LAM, 38 Rue Frédéric Joliot Curie, 13338 Marseille, France
        \and
        French-Chilean Laboratory for Astronomy, IRL 3386, CNRS and Universidad de Concepci\'{o}n, Departamento de Astronom\'{i}a, Barrio Universitario s/n, Concepci\'{o}n, Chile
        \and
        Institut de Recherche en Astrophysique et Planétologie (IRAP), Université de Toulouse, CNRS, UPS, CNES, Toulouse, France
        \and
        Valongo Observatory, Federal University of Rio de Janeiro, Ladeira do Pedro Ant\^{o}nio 43, Rio de Janeiro, RJ 20080-090, Brazil
        \and
        Instituto de Astrof\'{i}sica e Ci\^{e}ncias do Espa\c{c}o, Universidade do Porto – CAUP, Rua das Estrelas, PT4150-762 Porto, Portugal
        \and
        Departamento de F\'{i}sica e Astronomia, Faculdade de Ci\^{e}ncias, Universidade do Porto, Rua do Campo Alegre 687, PT4169-007 Porto, Portugal
        }
        
   \date{Received 7 July 2026 / Accepted 4 November 2026}
 
  \abstract
  {We study the resolved stellar kinematics and star formation histories, and their dependence on the environment, in a sample of 218 intermediate redshift (0.1 < z $\lesssim$ 1) galaxies located in the CANDELS and COSMOS fields, observed with VLT/MUSE. The sample covers galaxies with stellar masses ranging M$_{\star}$ $\sim$ 10$^{8}$ - 10$^{12}$ M$_{\odot}$, including both star-forming and quiescent systems. For 113 galaxies, found in dense environments of galaxy groups, we derive the spin parameter ($\lambda_{R}$), while for the rest of the sample, we use data from previous papers in this series. Given that the surveys used in this work are flux-limited, the $\lambda_{R}$ distribution of our full sample lacks the bimodality observed in the local Universe, but the subsample of galaxies at z < 0.6 resembles the galaxy population in the nearby Universe, in terms of mass, sizes, and stellar population properties, more than galaxies at z > 0.6. We classified the galaxies into fast and slow rotators, finding that $\sim$8$\%$ (17/218) of the galaxies are slow-rotating. The fraction of slow rotators does not change significantly down to z = 0, but we find that in the z $\sim$ 0 sample, the spin parameter is, on average, higher than at higher redshifts. In our subsample at z > 0.6, we observed evidence for the existence of the kinematic morphology-density relation. The slow rotators in our sample mostly assemble their stellar mass in one or two main episodes of formation, and they finished assembling their mass via star formation by redshift of z $\sim$ 1. It is however, not clear if the loss of angular momentum can be associated with the last star formation episode.  Based on their location in the mass - size diagram, the intermediate-redshift galaxies are consistent with being the progenitors of the most massive slow rotators observed at z $\sim$ 0, where the subsequent growth is facilitated through gas-poor (dry) mergers.}

   \titlerunning{Inter-z slow rotators as the progenitors of the most massive galaxies at z $\sim$ 0}
   \authorrunning{Constanza Mu\~{n}oz L\'{o}pez et al.}
   \maketitle
%
\section{Introduction}

In the hierarchical picture of galaxy mass assembly, two important parameters characterising the galaxy evolution are: the angular momentum, which affects the gas dissipation and varies during mergers, and the type of stellar population, which carries the fossil record of different events of star formation. Galaxies, in terms of kinematics, evolve from rotation-dominated to velocity-dispersion-dominated systems. Morphological changes, variations in their stellar mass content, and the evolution of their stellar populations are some of the phenomena that accompanied this kinematical transition. 

With the advent of integral field spectroscopy (IFS), the stellar kinematics of galaxies have been studied not only through $v$/$\sigma$ ratios \citep{1978ComAp...8...27B,2005MNRAS.363..937B}, but also using their two-dimensional spatial information. \cite{2011MNRAS.414..888E}, using early-type galaxies (ETGs) from the local survey ATLAS$^{\rm 3D}$ \citep{2011MNRAS.413..813C} classified galaxies into two categories, given their stellar angular momentum quantified by the proxy $\lambda_{\rm R}$ \citep{2007MNRAS.379..401E} and the observed galaxy's ellipticity: fast rotators (FRs), galaxies where the stellar rotation happens predominantly in stellar disks, where stars are moving around the disk plane in ordered orbits, dominating the kinematics of the system \citep{2008MNRAS.390...93K}. On the other hand, slow rotators (SRs) are galaxies with low stellar angular momentum. Typically, SRs are velocity dispersion-dominated systems, where stars belong to multiple orbital families \citep{1987ApJ...321..113S,2007MNRAS.379..418C,2008MNRAS.385..647V, 2026A&A...708A..15K}, while their shape is non-axisymmetric \citep{2014MNRAS.444.3340W,2017MNRAS.472..966F}. However, some SRs can present stellar disks, as is the case with counter-rotating galaxies, which are systems consisting of two stellar disks with respective angular momentum vectors pointing in opposite directions. The resulting effect is that the counterrotation diminishes the overall stellar angular momentum of the galaxy \citep{2007MNRAS.379..418C,2011MNRAS.414.2923K,2017MNRAS.464.4789M,2019A&A...623A..87N,2020A&A...635A.129K}.

Furthermore, slow- and fast-rotating galaxies also present differences in their stellar populations. Several spectroscopic nearby surveys (ATLAS$^{\rm 3D}$; MaNGA, \citealp{2015ApJ...798....7B}; SAMI, \citealp{2012MNRAS.421..872C}; MASSIVE \citealp{2014ApJ...795..158M}, etc.) have studied galaxies’ stellar properties. In the local Universe, passive galaxies with M$_{\star}$ $\leq$ 10$^{11}$M$_{\odot}$ are axisymmetric \citep{2011MNRAS.414.2923K}, present disc-like structure \citep{2011MNRAS.416.1680C}, and are rotation dominated \citep{2011MNRAS.414..888E,2013MNRAS.432.1709C}. Moving to higher redshifts, the LEGA-C survey \citep{2016ApJS..223...29V,2018ApJS..239...27S} has studied galaxies at z $\sim$ 0.7 using the information provided by long-slit spectroscopy. They found that at this redshift, almost all star-forming galaxies and half of the quiescent population are rotation-dominated \citep{2021ApJ...923...11V}. Hence, the majority of SRs are massive galaxies, mostly detected in dense environments, and with low levels of recent star formation. 

The properties of SRs indicate these systems as the final product of galaxy evolution. However, it is an open question when, in cosmic time, the population of SRs emerged in a statistically significant number. Various studies have quantified the stellar angular momentum of galaxies at redshifts z > 0. \cite{2024A&A...688A..75M,2026A&A...707C...4M}, used a sample of 106 intermediate-redshift galaxies (0.1 < z < 0.8), observed by the spectrograph in the Very Large Telescope (VLT), Multi Unit Spectroscopic Explorer (MUSE, \citealp{2010SPIE.7735E..08B}), detecting SRs at z > 0.6. They found three SRs in their sample, with only two of them at z > 0.6. \cite{2024MNRAS.531.4602D}, using the MAGPI survey of galaxies at z $\sim$ 0.3, found that the fraction of slow-rotating galaxies was similar to MaNGA (z $\sim $  0) samples, indicating that the population of massive SRs had already emerged $\sim$4 Gyr ago. \cite{2026ApJ..1009...87M} quantified the rotational support of galaxies at 0.25 < z < 0.75, finding that galaxies are systematically more rotationally supported at that redshift range than their local counterparts. 

At z > 1, the spin parameter ($\lambda_{R}$), a proxy for the angular momentum, has been computed for small samples of galaxies. \cite{2018ApJ...862..126N} measured the stellar kinematics of four quiescent galaxies, gravitationally lensed by galaxy clusters at z = 1.95-2.64, using the FIRE echellette spectrograph \citep{2013PASP..125..270S} on the Magellan Baade telescope. All of the galaxies in the sample were classified as fast rotators.
Since the launch of the James Webb Space Telescope (JWST), studies have reported measured spin parameters for massive quiescent galaxies using NIRSpec \citep{2022A&A...661A..80J}. \cite{2025A&A...702A.110S} using slit-based observations from NIRSpec/MSA showed that 10/15 galaxies at z $\sim$ 1.2 - 2.3 exhibit significant rotation and are classified as fast rotators. Various studies based on NIRSpec/IFU observations at z > 3 have also estimated the stellar angular momentum. \cite{2024NatAs...8.1443D} measured $\lambda_{Re}$ for the galaxy GS-10578 at z $\sim$ 3, finding that the system is rotation-supported. \cite{2026MNRAS.547ag210P}, also measured the stellar kinematics of the galaxy GS-9209 at z =4.66, finding clear signs of rotation. \cite{2026NatAs..10.1232F} studied the kinematics of three massive quenched galaxies at z $\sim$ 3.5. They found that the galaxy, XMM-VID1-2075, is near the threshold separating the slow and fast rotators, but formally is classified as a fast rotator.
These massive quiescent galaxies at z $\sim$ 2-4 are the natural targets for finding the advent of SRs.

The kinematic morphology–density relation \citep{2011MNRAS.416.1680C}, analogous to the morphology–density relation \citep{1974ApJ...194....1O,1976ApJ...208...13D,1980ApJ...236..351D} but based on the stellar kinematics properties of galaxies, suggests that SRs at z $\sim$ 0 prefer populating high-density region, suggesting that they are formed more efficiently in denser environments, specifically as the central galaxies of massive clusters.

The role that the environment plays in the formation and evolution of galaxies has been the focus of several studies \citep{2011MNRAS.414..888E,2011MNRAS.416.1680C,2021MNRAS.508.2307V,2017ApJ...844...59B,2019A&A...632A..59F,2019191005139G,2024arXiv240206877C,2024MNRAS.528.5852V} since the discovery of the kinematic morphology–density relation. It remains an open question how, or whether, changes in galaxies' morphology affect their structural and kinematical properties. For this reason, it is key to have large samples that cover the wide diversity of galaxies across various environments and at different cosmic times. 

Various cosmological simulations have studied when the slow-rotating population emerges. \cite{2018MNRAS.480.4636S}, using the cosmological hydrodynamical Magneticum Pathfinder simulations, investigated the fast and slow-rotating populations, finding that the latter starts building up at z $\sim$ 2, and gets statistically significant at z $\sim$ 1. \cite{2025ApJ...979...15K} observed the formation of a slow-rotating population emerging at z > 4 and increasing in number towards lower redshifts, also using the Magneticum Pathfinder simulation. 

The path for fast-rotating galaxies to decrease their angular momentum and become velocity dispersion-dominated systems remains debatable. Various simulations have shown that mergers are processes that affect galaxies' stellar kinematics \citep{2009A&A...501L...9D,2011MNRAS.416.1654B,2014MNRAS.444.3357N,2018MNRAS.480.4636S}. Moreover, some studies suggest that gas-poor or dry mergers have the effect of decreasing $\lambda_{\rm R}$ \citep{2009ApJ...705..920H,2010ApJ...723..818H,2014MNRAS.445.1065R,2014MNRAS.444.3357N,2018MNRAS.476.4327L}. \cite{2018MNRAS.476.4327L}, using the hydrodynamical simulations EAGLE and HYDRANGEA, observed that $\sim$70$\%$ of the SRs in the local Universe have experienced at least one important merger (with mass ratio $\geq$ 0.1), with gas-poor mergers occurring more commonly than wet mergers. \cite{2022MNRAS.509.4372L}, using an EAGLE sample of z = 0 galaxies, found that the SRs that do not experience mergers during their lifetime have different intrinsic properties than SRs that have had major or minor mergers. Non-merger SRs have smaller sizes (more compact), present lower black hole-to-stellar mass ratios than other SRs at fixed stellar mass, reach quiescence later, and contain specific characteristics in their stellar populations (lower [$\alpha$/Fe] and higher [Fe/H] ratios than other SRs).
\cite{2024MNRAS.535.1202R} studied the impact of major and minor megers on the formation of cores, the size growth, and on the stellar kinematics of galaxies at z $\sim$ 2-3. The presence of supermassive black holes is related to the formation of low-density cores by displacing stars out of the central region of galaxies which experienced a major merger. The displaced stars contribute to the growth of galaxies in terms of their size (as measured via their half-mass radius). In addition, some of such mergers ($\sim$6-8 minor mergers) are also capable of transforming a FR into a SR.

In this paper, we combine the results on the stellar kinematics from 
\cite{2024A&A...688A..75M} and stellar populations of \cite{2025A&A...702A.146M}, with a new set of stellar kinematics of galaxies from the MUSE/MAGIC survey \citep{2024A&A...683A.205E}. Our aim is to look at the connection between the SFH and the spin parameter and investigate the properties of SRs. The paper is structured as follows. In Section \ref{sec:sample_paper3}, we present the data set and galaxy sample. In Section \ref{sec:method_paper3}, we describe the method used to build the sample. In Section \ref{sec:results_paper3}, we present the main results and discuss their interpretation. In Section \ref{sec:conclusions_paper3}, we present the main conclusions of our work. Throughout the paper we adopt a flat Universe, H$_{0}$ = 70 km s$^{-1}$ Mpc$^{-1}$, $\Omega_{\Lambda}$ = 0.7 cosmology.

\section{Data sets and galaxy sample}\label{sec:sample_paper3}

This work relies on data taken from four main spectroscopic surveys. The spectroscopic surveys are located in the CANDELS/GOODS-S and CANDELS/COSMOS regions. MUSE-Wide \citep{2019A&A...624A.141U} covers $100 \times 1$ arcmin$^{2}$ fields each observed for one hour, while MUSE-Deep \citep{2017A&A...608A...1B, 2023A&A...670A...4B} covers 9 x 1 arcmin$^{2}$ mosaic for 10 hours, $1 \times 1$ arcmin$^{2}$ field for 31 hours, and 141 hours on one circular field with $1^{\prime}$ diameter \citep[MXDF;][]{2023A&A...670A...4B}, in the Hubble Ultra Deep Field \citep[HUDF;][]{2006AJ....132.1729B} region. Additionally, the MAGIC survey \citep{2024A&A...683A.205E} targets 14 massive galaxy structures, mostly groups, chosen from the zCOSMOS 20k catalogue \citep{2009ApJ...697.1842K,2012ApJ...753..121K}, having exposure times from 1 to 10 h. 

MUSE-Wide has a wide area of coverage and shallower data, given its integration time. On the other hand, MUSE-Deep and MXDF reach lower masses and higher redshift galaxies, given their high exposure times. The MAGIC survey probes different environments (overall more dense than galaxies in \citealp{2024A&A...688A..75M}), and increases the number of massive galaxies at higher redshift, z $\sim$ 0.7, where it is mainly targeting dense structures. The galaxies in the sample located within the footprint of MUSE-Wide, MUSE-Deep, and MXDF are presented in \cite{2024A&A...688A..75M,2026A&A...707C...4M}. These galaxies correspond to Galaxy 1 to Galaxy 105 in our sample (see Table \ref{tab:sample_table}). The remaining galaxies in the sample (from Galaxy 106 to 218) are located within the MAGIC footprint.

    \begin{figure}
    \centering
       \includegraphics[width=7.8cm]{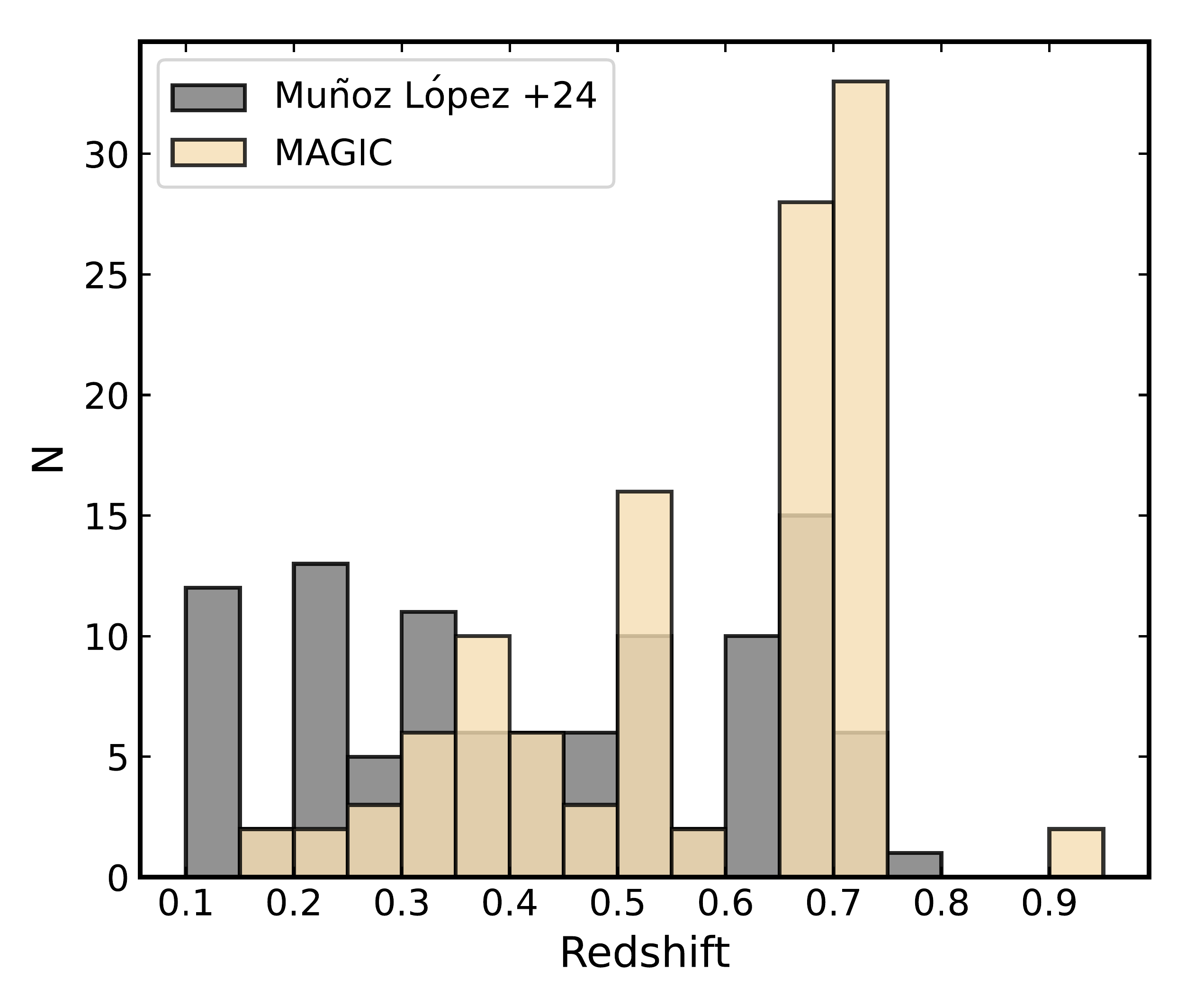}
         \caption{Redshift distribution of the galaxy sample. Number of systems as a function of the galaxy's redshift. Grey bars show the sample galaxies from \cite{2024A&A...688A..75M}. Wheat-coloured bars show newly added galaxies from the MAGIC survey \citep{2024A&A...683A.205E}.}
         \label{fig:red_distr_paper3}
    \end{figure}

\subsection{Selection criteria}\label{sec:selection_paper3} 

To increase and homogenise the spectral S/N of the sample galaxies, we spatially binned them using a code based on Voronoi tessellation developed by \cite{2003MNRAS.342..345C}. We required, for each galaxy, that they are represented by at least five Voronoi bins, each of them with S/N $\geq$ 8 in the rest-frame wavelength range 4150-4350 $\AA$, which is representative of the continuum level. 

Therefore, the sample is composed of 218 galaxies. In \cite{2024A&A...688A..75M}, we presented stellar kinematics of 104 galaxies included in our current sample. In this paper, we present the unpublished stellar kinematics for 113 galaxies in the MAGIC survey and one additional system not included in \cite{2024A&A...688A..75M}.

\subsection{Global properties}\label{sec:global-properties_paper3}

The sample is composed by 218 galaxies: 105 galaxies within the MUSE-Wide footprint (i.e. MUSE-Wide, HUDF, MUSE-Deep and MXDF) presented in \cite{2024A&A...688A..75M} (with the exception of Galaxy 105), with stellar masses ranging from $\sim$10$^{7.8}$M$_{\odot}$ to 10$^{12}$M$_{\odot}$, and 113 galaxies members of the MAGIC survey spanning from $\sim$ 10$^{9.3}$M$_{\odot}$ to 10$^{11.3}$M$_{\odot}$. The sample covers low- to intermediate-density environments and small to large groups (MAGIC footprint).
 
In Fig. \ref{fig:red_distr_paper3}, we present the redshift distribution of the sample. Our sample ranges from z $\sim$ 0 to z $\sim$ 1, covering more than 6 Gyr in cosmic time. We see that most of the MAGIC galaxies are at z $\sim$ 0.7, while galaxies from \cite{2024A&A...688A..75M} are covering the redshift range more evenly between z $\sim$ 0.1 and z $\sim$ 0.8. 

    \begin{figure}
    \centering
       \includegraphics[width=9.cm]{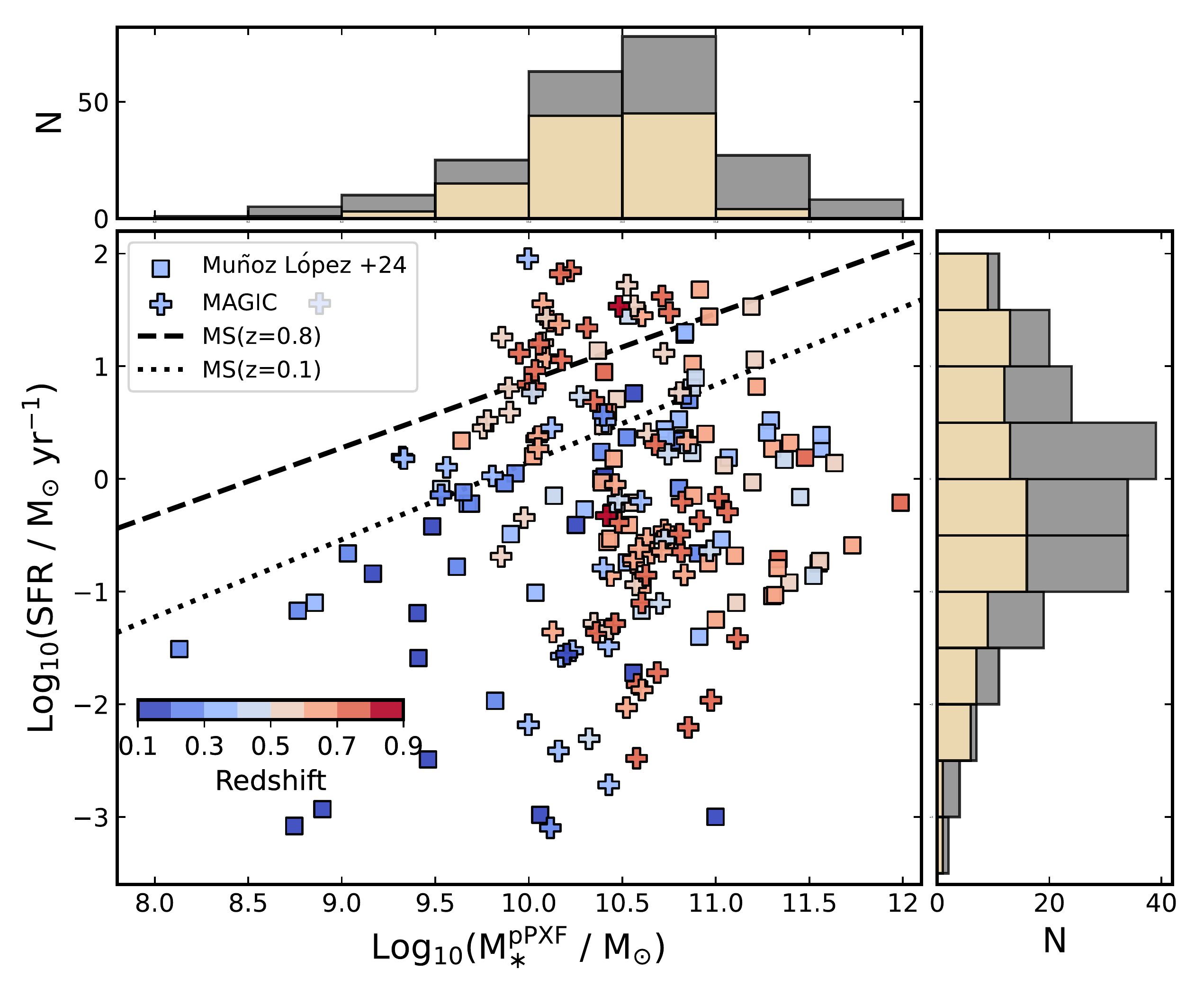}
         \caption{General properties of the galaxy sample. Star formation rate as a function of the stellar mass (derived from the MUSE spectra) for the galaxy sample, coloured coded by the spectroscopic redshift. The dotted and dashed black lines indicate the main sequence (MS) for an age of the Universe respectively of 12.4 Gyr (z = 0.1) and 6.8 Gyr (z = 0.8), given by Eq. (1) of \cite{2017ApJ...838...19W}. The stellar mass and SFR distributions are shown by the histograms. Grey bars show the sample galaxies from \cite{2024A&A...688A..75M}. Wheat-coloured bars show galaxies from the MAGIC survey \citep{2024A&A...683A.205E}.}
         \label{fig:sfr_mass_red_paper3}
    \end{figure}

In Fig. \ref{fig:sfr_mass_red_paper3}, we present the star formation rates (SFR) as a function of the stellar mass for the sample galaxies. The SFR for galaxies within the MUSE-Wide area was determined using the FAST code \citep{2009ApJ...700..221K} as part of the 3D-HST program \citep{2014ApJS..214...24S}. In the case of MAGIC galaxies, the SFR was computed using CIGALE code \citep{2019A&A...622A.103B} within an aperture of 3'' of diameter for each galaxy, and presented in \cite{2024A&A...683A.205E}. This aperture contains most of the galaxy light. Overall, our sample covers galaxies with stellar masses ranging from $\sim$10$^{8}$M$_{\odot}$ to 10$^{12}$M$_{\odot}$, and contains from star-forming to quenched systems. 
We also have low-mass systems (M$_{\star} \leq$ 10$^{9.5}$ M$_{\odot}$), which are galaxies at low redshift and mostly below the star-forming sequence. Thus, our sample lacks galaxies with low masses \citep{2018A&A...619A..27B} and high levels of star formation. Generally, low-mass star-forming galaxies are populating low-density environment  \citep{1988ApJ...330L..17I, 2000MNRAS.311..307T, 2014MNRAS.445.1694L} and usually present low surface brightness \citep{1993AJ....106.1784C, 1995MNRAS.275....1T}. Hence, they are not bright and massive enough to be included in the sample. In Appendix \ref{sec:sample_table_paper3}, Table \ref{tab:sample_table}, we list the properties of each galaxy in the sample. 

\section{Method}\label{sec:method_paper3}

\subsection{Stellar kinematics}\label{sec:kin_paper3}

We used the pPXF algorithm for extracting the resolved stellar kinematics of the galaxy sample. We followed a similar approach presented in \cite{2024A&A...688A..75M}. We recovered the first and the second-order moments (the radial mean velocity V, and the velocity dispersion $\sigma$) of the line-of-sight velocity distribution (LOSVD), given the spatial resolution and signal-to-noise ratio (S/N) of our sample (each galaxy has an integrated S/N $\geq$ 10 $\AA^{-1}$).

For the kinematic fitting, we used 195 templates from the Indo-US stellar library \citep{2004ApJS..152..251V} due to their finer and constant spectral resolution of 1.35 $\AA$ full width at half maximum (FWHM), which is better than the MUSE line-spread function (LSF) FWHM in all its wavelength range.

As we wanted to extract the stellar kinematics, we masked several gas emission lines: [OII] $\lambda \lambda$3726,3728, H$\delta$ $\lambda$4101, H$\gamma$ $\lambda$4340, H$\beta$ $\lambda$4861, [OIII] $\lambda \lambda$4958,5006, [OI] $\lambda$6300, [NII] $\lambda$6548, 6583, H$\alpha$ $\lambda$6563, and [SII] $\lambda \lambda$6716, 6730. As some of our galaxies were observed by MUSE using adaptive optics (AO), we masked the nominal wavelength range from 480-930 nm when pertinent. For the spectral fitting, we adopted additive polynomials of 12th order and a zero-degree multiplicative polynomial.

Before the fitting started, we convolved the stellar templates to a wavelength-dependent MUSE LSF resolution as derived in \cite{2017A&A...608A...5G}. We spatially binned the galaxy sample using a code based on Voronoi tessellation developed by \cite{2003MNRAS.342..345C}. We produced kinematic maps where each bin has at least a integrated S/N $\sim$ 8 in the wavelength range of 4150-4350 $\AA$. This wavelength range is an interval free of emission lines and with weak absorption features, therefore representative of the continuum level. All the galaxies in our sample have at least 5 Voronoi bins.

For each bin, we computed the errors in velocity and velocity dispersion by performing a bootstrap approach. We added random Gaussian noise to the fitted spectrum of each bin and fit the new spectrum with pPXF again. We created a total of 100 spectra, and from the extracted mean velocity and velocity dispersion, we estimated the errors in V and $\sigma$ of the original/observed bin spectrum associated to that bin. We present the kinematic maps of our the galaxy sample in the Appendix \ref{sec:kin_maps_appendix_paper3}. 

We quantify the global rotational velocity structure of galaxies using the kinematic maps by computing the spin parameter $\lambda_{R}$, a proxy for the luminosity-weighted stellar angular momentum. $\lambda_{R}$ is calculated using equations (5) and (6) of \cite{2007MNRAS.379..401E}:

    \begin{equation}
    \lambda_{R} \equiv \frac{\langle R \mid V \mid \rangle}{\langle R \sqrt{V^{2} +\sigma^{2}}\rangle} = \frac{\sum_{i=1}^{N} F_{i} R_{i} \mid V_{i} \mid}{\sum_{i=1}^{N} F_{i} R_{i} \sqrt{V^{2}_{i} +\sigma^{2}_{i}}}
    \label{eq:lam}
    .\end{equation} 
Where $F_{i}$, $R_{i}$, $V_{i}$, and $\sigma_{i}$ are the flux, the distance to the centre of the galaxy, the mean stellar velocity after removing the systemic velocity, and the velocity dispersion of the $i$-th pixel, respectively. The centre of the galaxy is determined to be the pixel with the highest integrated flux. The values for $V$ and $\sigma$ for a bin are calculated with the flux contribution of each pixel inside it. Thus, the individual $V_{i}$ and $\sigma_{i}$ values for a particular pixel correspond to the bin velocity and velocity dispersion. Moving to the original pixel space, the value for $F_{i}$ corresponds to the original flux of the pixel ($F_{i}$). In this way, each pixel keeps its own flux, whereas its velocity and velocity dispersion are inherited from the bin it belongs to. $\lambda_{R}$ is normalized by $\sqrt{V^{2}_{i} +\sigma^{2}_{i}}$, which is proportional to stellar mass, implying that the spin parameter goes to unity when the mean stellar velocity ($V$) dominates. The summation is performed over $N$ pixels within a fixed aperture. For our sample, we compute the spin parameter between an elliptical area that better covers the available galaxy's data (within 1-2R$_{e}$). 

As the atmospheric seeing blurs the kinematics of galaxies, we corrected the spin parameter, following \cite{2018MNRAS.477.4711G} prescription, given by:
\begin{equation}
        \lambda_{{\rm R}}^{{\rm obs}} =   \lambda_{{\rm R}}^{{\rm true}} gM_{2} \left ( \frac{\sigma_{{\rm PSF}}}{R^{\rm maj}} \right ) f_{n} \left ( \frac{\sigma_{{\rm PSF}}}{R^{\rm maj}} \right )
    \label{eq:correc}
    ,\end{equation} 

where $\lambda_{R}^{{\rm obs}}$ and $\lambda_{R}^{{\rm true}}$ are the observed (or measured) and the intrinsic (or true) angular momentum, respectively; $R^{{\rm maj}}$ is the semi-major axis of the aperture within which $\lambda_{R}$ is calculated; $n$ is the S\'{e}rsic index; and $\sigma_{{\rm PSF}}$ = ${\rm FWHM_{PSF}}/2.355$; $f_{n}$ is an empirical function used to introduce the S\'{e}rsic index dependence into the correction; and $gM_{2}$ is the generalized Moffat function \citep{1969A&A.....3..455M}. The correction is only appropriate for galaxies with an angular size larger than the PSF (FWHM) (i.e. $\sigma_{{\rm PSF}} \leq R^{{\rm maj}}$). For the newly added galaxies from the MAGIC survey, we use the values for the PSF (FWHM) provided in Table 1 of \cite{2024A&A...683A.205E}.
    \begin{figure}
    \centering
       \includegraphics[width=9.cm]{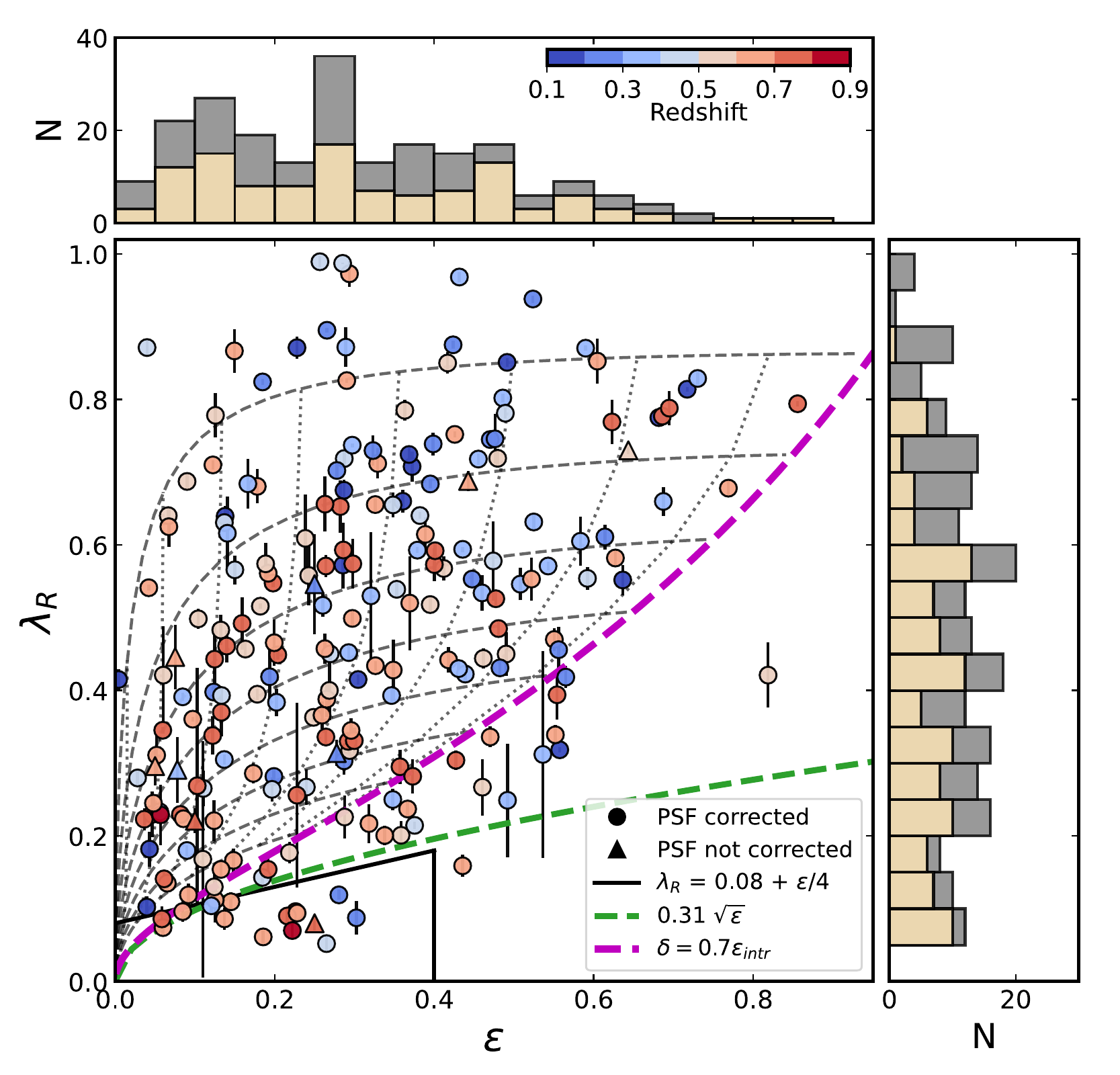}
         \caption{$\lambda_{R}$ as a function of the galaxy's ellipticity coloured by the redshift for the 218 galaxies in the sample. Circles are systems where the spin parameter has been corrected by the atmospheric seeing. Triangles are galaxies were the PSF correction has not been applied, thus they represent lower limit values. The green dashed line denotes the threshold between slow and fast rotators, given by 0.31 $\times$ $\sqrt{\epsilon}$ \citep{2011MNRAS.414..888E}. The black trapezium defines the region for slow rotator ETGs: $\lambda_{R}$ < 0.08 + $\epsilon$/4, $\epsilon$  < 0.4 \citep{2016ARA&A..54..597C}. The purple line shows the edge-on anisotropy-ellipticity relation $\delta$ = 0.7 $\times$ $\epsilon_{\rm intr}$ from \cite{2007MNRAS.379..418C}. The thin dotted lines correspond to the magenta line at different inclinations. The black dashed lines trace how galaxies with a particular $\epsilon_{\rm intr}$ would move with changing inclination, using the relations from \cite{2016ARA&A..54..597C}. The galaxy's ellipticity and spin parameter distributions are shown by the histograms. Grey bars show the sample galaxies from \cite{2024A&A...688A..75M,2026A&A...707C...4M}. Wheat-coloured bars show galaxies from the MAGIC survey \citep{2024A&A...683A.205E}.}
         \label{fig:eps_lam_correc_paper3}
    \end{figure}

\subsection{Stellar populations}\label{sec:star_pop_paper3}

We used the stellar population parameter and star formation histories derived in \cite{2025A&A...702A.146M} and added an additional 30 galaxies in this work.

Full details regarding the methods for deriving the stellar population parameters, stellar masses, star formation history, and star formation rates methods are presented in sections 3.1, 3.2, 3.3, and 3.4 of \cite{2025A&A...702A.146M}, respectively.

We extracted the stellar population parameters from the absorption-line spectra of the sample galaxies within 1R$_{\rm e}$ using pPXF, and the stellar population library E-MILES \citep{2006MNRAS.371..703S, 2011A&A...532A..95F}. By using pPXF together with photometry measurements in different bands, we retrieve the stellar masses of the galaxies. From the resulting fitting procedures, we derived the mean stellar ages, metallicities, star formation histories, and number of star-forming episodes for the galaxy sample.

\section{Results and discussion}\label{sec:results_paper3}

\subsection{Spin parameter}\label{sec:spin_paper3}
 
The corrected spin parameter, computed here for 113 galaxies and taken from \cite{2024A&A...688A..75M,2026A&A...707C...4M}, for the rest, is presented in Fig. \ref{fig:eps_lam_correc_paper3}. Figure \ref{fig:spin_obs_paper3} shows the observed spin parameter (i.e. before PSF correction). Out of 218, nine galaxies can not be PSF-corrected, which are shown by triangles in Fig. \ref{fig:eps_lam_correc_paper3}, and correspond to lower limits for the stellar angular momentum. The correction cannot be implemented when the resolution parameter ($r$ = $\sigma_{\rm{PSF}}/ \rm{R^{maj}}$) is $r$ > 1, and when the galaxy spin parameter correction is not physical (i.e., $\lambda_{R}^{\rm true} >1$). We observe that our sample is dominated by round ($\epsilon$ < 0.3) galaxies and systems with $\lambda_{R}$ < 0.8. Our sample also has a low density of galaxies with high values of spin parameter, in the full ellipticity range. We do not observe any trend of the spin parameter with increasing redshift in our sample. We have a total of 17 slow rotators in our sample, corresponding to $\sim$8$\%$ of the total number of galaxies in the sample. We found a higher fraction of slow rotators in denser environments: among the 113 galaxies within the MAGIC footprint, $\sim$12$\%$ (14/113) are SRs. On the other hand, $\sim$3$\%$ (3/105) of the sample galaxies within the MUSE-Wide footprint are SRs.

\begin{figure}
    \centering
       \includegraphics[width=8.cm]{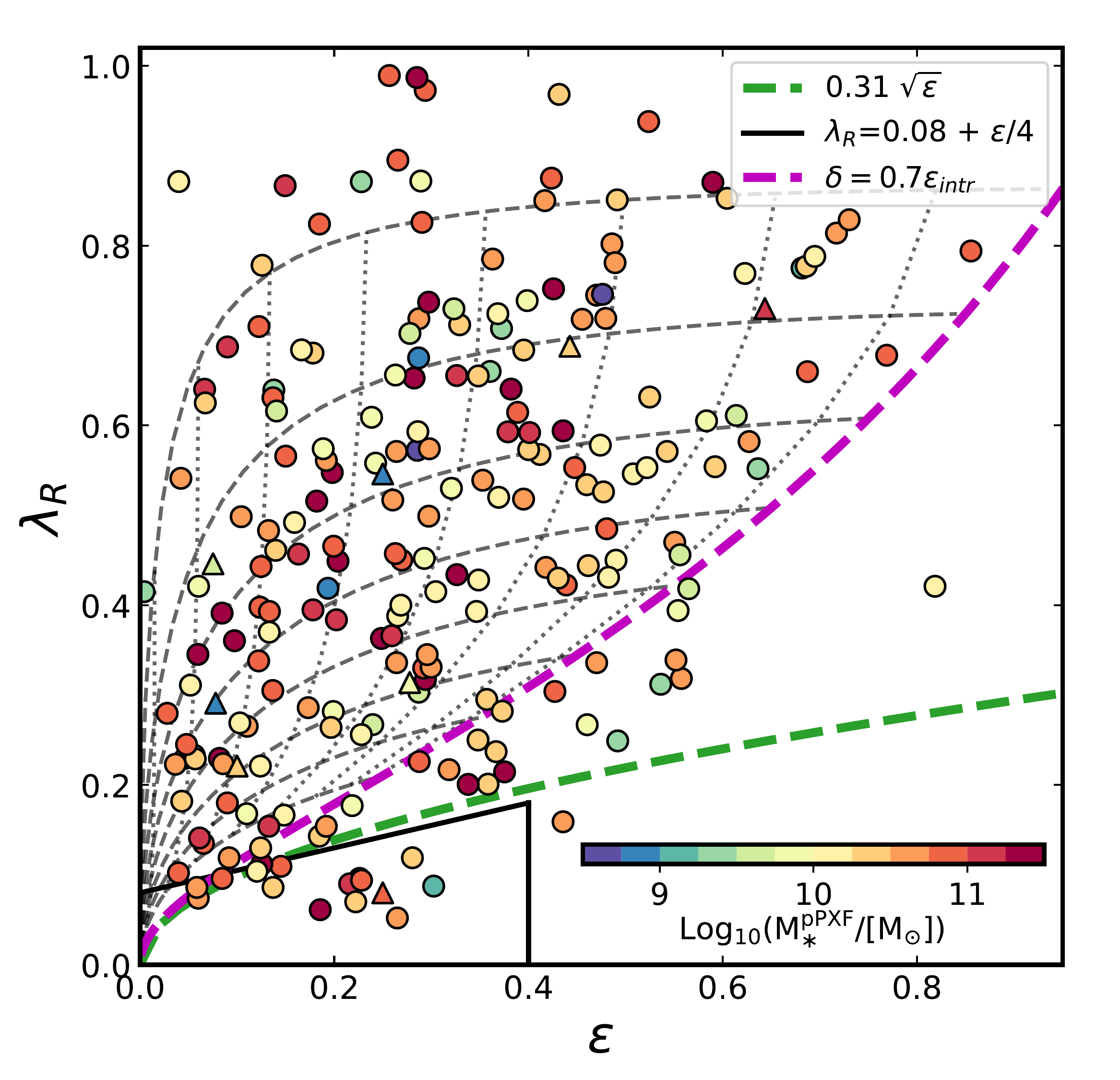}
         \caption{$\lambda_{R}$ as a function of the galaxy's ellipticity coloured-coded by the stellar mass for the 218 galaxies in the sample. Full description of symbols and lines is given in Fig. \ref{fig:eps_lam_correc_paper3}.}
         \label{fig:eps_lam_mass_paper3}
    \end{figure}

\cite{2018MNRAS.477.4711G} studied the stellar angular momentum, using a MaNGA sample of $\sim$2300 galaxies in the nearby Universe. In their sample they observed a bimodality in the $\lambda_{R}$ distribution, peaking around $\lambda_{R}$ $\sim$ 0.9 and 0.1. We do not observe this bimodality in our sample, but record a broad distribution of $\lambda_{R}$ centered around 0.5. The number of galaxies drops significantly for $\lambda_{R}$ > 0.7. Below $\lambda_{R} \sim$ 0.15, where slow rotators are expected to be found, there is a notable decrease in the number of galaxies. As $\lambda_{R}$ further decreases, there is a minor increase in galaxies, signifying an increase in the population of velocity dispersion dominated systems. In Fig. \ref{fig:eps_lam_mass_paper3} we show the $\lambda_{R}$-$\epsilon$ diagram coloured by the stellar mass. At $\lambda_{R}$ > 0.8, the diagram is sparsely populated, and the galaxies in this spin range are mostly more massive than M$_{\star}$ > 10$^{10.5}$ M$_{\odot}$. These characteristics of our sample differs from \cite{2018MNRAS.477.4711G} sample, where they have a high fraction of low mass galaxies at high angular momentum. At z = 0, the high spin parameter ($\lambda_{R}$ > 0.8) is dominated by low-mass (M$_{\star}$ < 10$^{10}$ M$_{\odot}$) galaxies. In Fig. \ref{fig:eps_lam_loess_paper3}, we present the LOESS (LOESS stand for locally weighted regression robust technique) smoothed version of Fig. \ref{fig:eps_lam_mass_paper3}, where we see more clearly that the mass trend observed in the MaNGA sample (Fig. 10 in \citealp{2018MNRAS.477.4711G}) is not seen in our $\lambda_{R}$ distribution. Therefore, the lack of a spin parameter bimodality in our sample is a result of an overall lack of lower mass galaxies, which is related to our sample being biased towards more massive systems (see more in Sect. \ref{sec:bias_paper3}).

\begin{figure}[hbt!]
\centering
   \includegraphics[width=8.6cm]{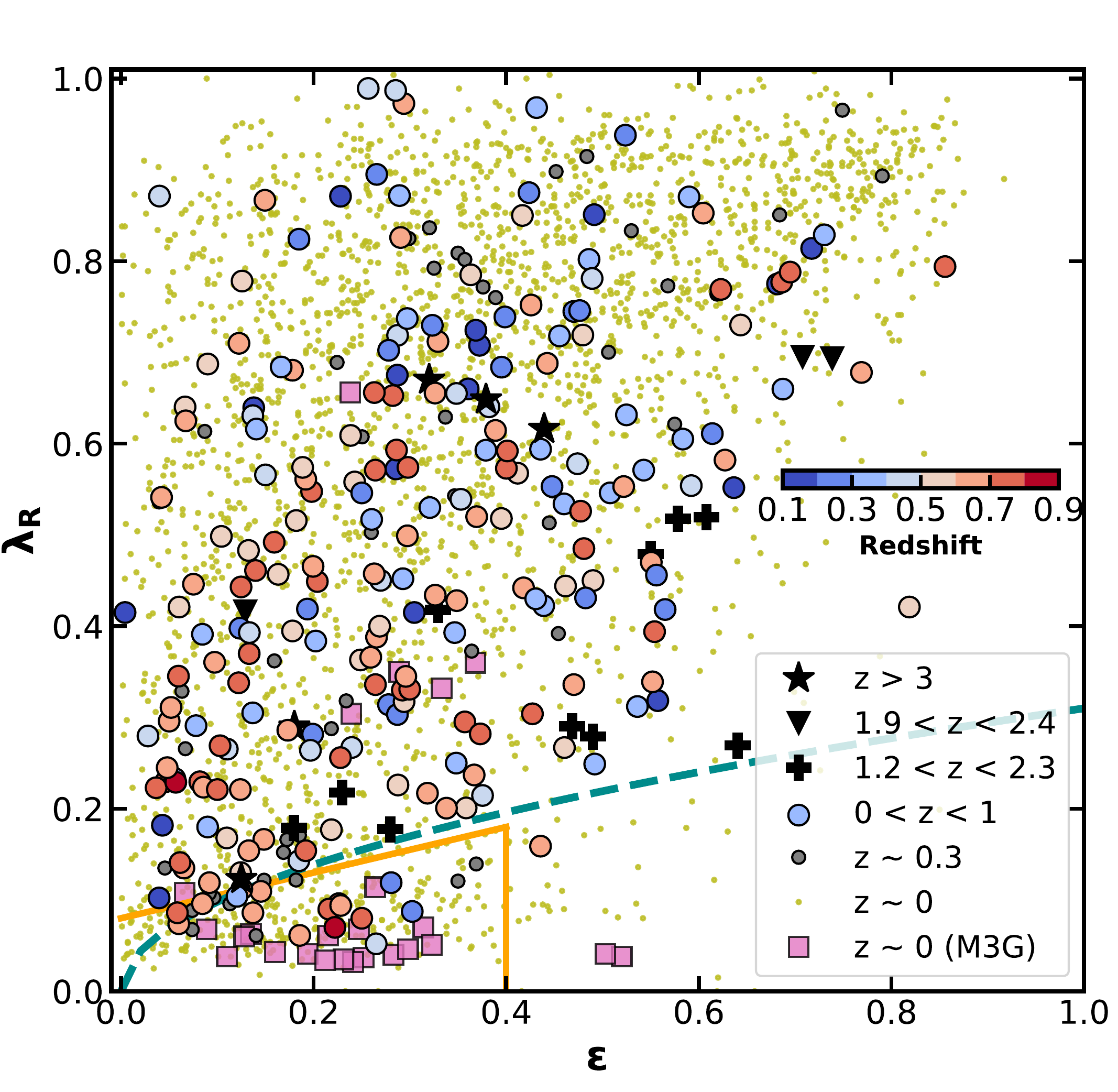}
      \caption{Stellar angular momentum of intermediate-redshift galaxies. $\lambda_{R}$ as a function of the galaxy's ellipticity coloured by the redshift of our sample (large filled circles). The teal dashed line denotes the threshold between slow and fast rotators, given by 0.31 $\times$ $\sqrt{\epsilon}$ \citep{2011MNRAS.414..888E}. The orange trapezium defines the region for SR ETGs: $\lambda_{R}$ < 0.08 + $\epsilon$/4, $\epsilon$  < 0.4 \citep{2016ARA&A..54..597C}. Pink squares correspond to galaxies from the MUSE Most Massive Galaxies (M3G) in the nearby Universe \citep{2026A&A...708A..15K}. Olive dots show galaxies at z $\sim$ 0 from the MaNGA survey, presented in \cite{2018MNRAS.477.4711G}. Grey circles are galaxies from the MAGPI survey at z $\sim$ 0.3 \citep{2024MNRAS.531.4602D}. Black symbols correspond to massive galaxies at z > 1 \citep{2018ApJ...862..126N,2024NatAs...8.1443D,2025A&A...702A.110S,2026NatAs..10.1232F,2026MNRAS.547ag210P}.}
              \label{fig:eps_lam_paper3}
\end{figure}

    \begin{figure*}
    \centering
       \includegraphics[width=18.5cm]{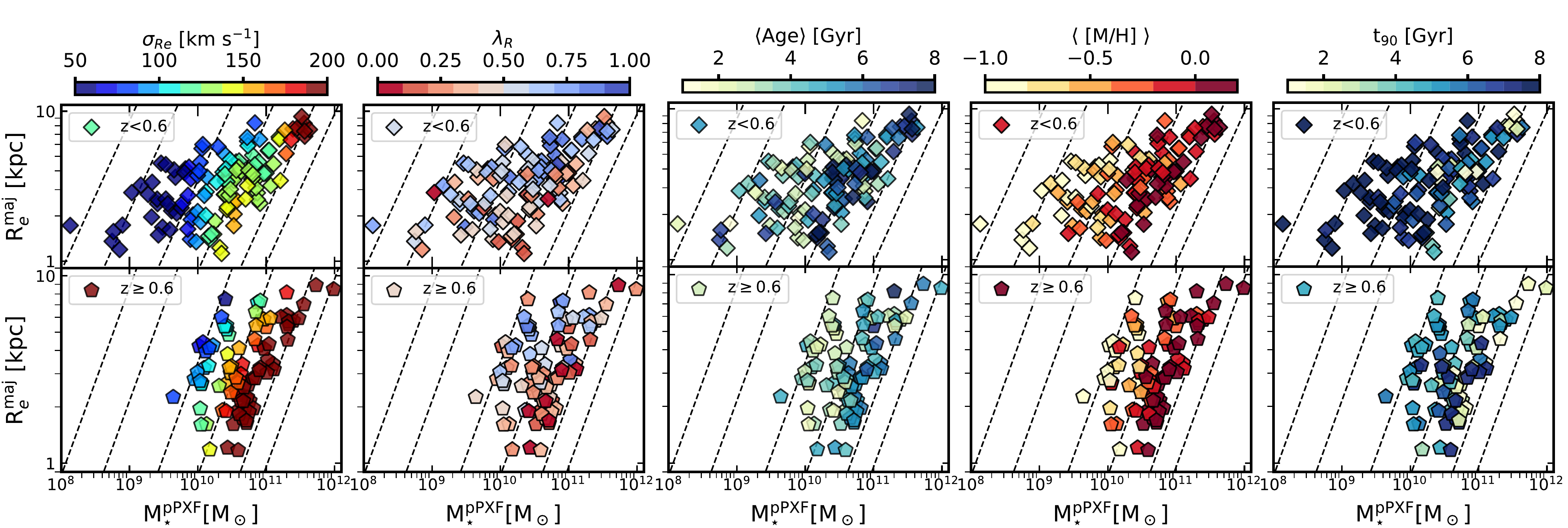}
          \caption{Plane of effective radius (R$^{\rm maj}_{\rm e}$) versus stellar mass coloured by stellar kinematics indicators and SSP-equivalent population parameters measured within one effective radius. Colours from the left to right panels correspond to the velocity dispersion, spin parameter, mean stellar age, mean stellar metallicity, and the time taken to form 90$\%$ of the stars in the galaxy, respectively. The velocity dispersion was spatially averaged using the LOESS technique \citep{Cleveland1988LocallyWR} implemented by \cite{2013MNRAS.432.1862C}. Dashed-black lines show constant velocity dispersion of 10, 20, 50, 100, 200, 300, km s$^{-1}$ from left to right, derived by the virial mass estimator M = 5R$^{\rm maj}_{\rm e}$ $\sigma^{2}$/G. Upper and lower panels corresponds to galaxies in the sample with redshift z < 0.6 and z > 0.6, respectively.}
         \label{fig:mass_size_paper3}
    \end{figure*}

In Fig. \ref{fig:eps_lam_paper3}, we show a summary of the spin parameter as a function of the observed ellipticity for galaxies at intermediate redshifts, collected from our sample and the sample of \cite{2024MNRAS.531.4602D}. We overplot values for z = 0 galaxies coming from the MaNGA \citep{2018MNRAS.477.4711G} and the M3G sample \citep{2026A&A...708A..15K} of the most massive z = 0 galaxies. In addition, we show all known $\lambda_{R}$ for z > 1 galaxies, based on JWST data \citep{2018ApJ...862..126N,2024NatAs...8.1443D,2025A&A...702A.110S,2026NatAs..10.1232F,2026MNRAS.547ag210P}. This figure highlights some of the points we will discuss further: 
\begin{itemize}
\item SRs are established as a statistically significant population by z $\sim$ 0.7, but there is little evidence for SRs in earlier epochs
\item  nearby massive galaxies have lower $\lambda_{R}$ than intermediate-redshift SRs.
\item Currently, the earliest known SR is at z = 0.943 (Galaxy 198)\footnote{\cite{2026NatAs..10.1232F}  found a dispersion-dominated galaxy at z $\sim$ 3 near the threshold separating the slow- from fast-rotating population. Nevertheless, the system formally corresponds to a fast rotator.}.
\end{itemize}

\subsection{Stellar population and kinematics}\label{sec:pop_kin_paper3}

In addition to the sample presented in \cite{2025A&A...702A.146M}, the stellar population parameters were extracted within one effective radius (R$_{\rm e}$) for $\sim$30 galaxies (within the MAGIC footprint), following the same recipe. The size parameter R$_{\rm e}^{\rm maj}$ corresponds to the semi-major axis of the half-light ellipse, that is defined as an ellipse covering the same area as a circle with radius R$_{e}$. 

Kinematic maps, according to their morphology, can be classified into regular (RR) and non-regular rotators (NRRs) \citep{2011MNRAS.414.2923K}. The classification quantifies how similar the galaxy's velocity map is to the velocity map of a stellar disc \citep{2008MNRAS.390...93K}. We use the software \texttt{KINEMETRY} \citep{2006MNRAS.366..787K}, which performs a harmonic analysis of velocity maps, differentiating between the disc-like rotation and complex velocity maps. We follow the same criteria for recognizing the regular rotation as done in \citet{2024A&A...688A..75M}. Therefore, we classified the kinematics of our sample into regular rotators (RR), non-regular rotators (NRR), galaxies with signs of regular and non-regular rotation but highly affected by PSF effects, as RR*, and NRR*, respectively. Our sample also contains galaxies with kinematic maps where the data quality is not sufficient for this type of analysis; thus, their maps are classified as uncertain (UN).

We find that $\sim$40$\%$ of our galaxies exhibit clear stellar rotation (RR), while $\sim$26$\%$ are classified as RR$^{*}$. Our sample also contains NRRs, around 19$\%$, while $\sim$9$\%$ of the kinematic maps are classified as NRR$^{*}$. The classification is not possible for $\sim$7$\%$ of systems; therefore, they belong to the UN class. Thus, $\sim$65$\%$ and $\sim$30$\%$ of the maps present RR and NRR features, respectively. Then our inter-z sample has a slight excess of NRR compared to the local samples, but it is not far off, and the differences might be due to the uncertainties (which will be discussed later).

We divide our galaxy sample into two redshift bins: high-redshift galaxies or galaxies at redshift z > 0.6, and low-redshift galaxies or systems at redshift z < 0.6. The sample contains more galaxies in the lower redshift bin. Out of the 218 galaxies in the sample, 123 are at z < 0.6, and 95 systems are at z > 0.6. Choosing a similar redshift separator, for example z = 0.5, does not change the results, but the imbalance between the sample sizes increases. In Fig. \ref{fig:mass_size_paper3}, we present the stellar mass-size relation, coloured by different population parameters, as well as the spin parameter, separated into two redshift bins. We observe that the higher redshift bin contains more massive galaxies (M$_{\star}$ > 10$^{10}$ M$_{\odot}$) than the lower redshift bin. These systems are characterised by shorter formation timescales t$_{90}$ as is expected given that for z > 0.6 the Universe is younger than 7.9 Gyr. Both higher and lower redshift bins contain galaxies ranging in size from 1 to 10 kpc.  Galaxies at higher redshift have higher values of velocity dispersion compared to their lower redshift counterparts, which is not surprising given that they are on average more massive. On the other hand, the least massive systems in our sample are galaxies with R$_{\rm e}^{\rm maj}$ < 4 kpc, found exclusively at the lowest redshifts (z < 0.4), and have the longest formation timescales and lowest mean metallicities in the sample.

We also observe that the parameters describing the stellar population are almost constant along lines of constant velocity dispersion (dashed lines in Fig. \ref{fig:mass_size_paper3}), derived from the virial estimator. This behavior is similar to galaxies at z $\sim$ 0 (e.g \citealp{2016ARA&A..54..597C}, Fig. 22). Given this characteristic along lines of constant $\sigma$, we observe that neither the physical size nor the stellar mass are individually the main drivers of the stellar properties in the galaxy sample.

If we look at the mass-size relation coloured by $\lambda_{R}$, shown in the second column of Fig. \ref{fig:mass_size_paper3}, we observe two main characteristics regarding the distribution of galaxies at high and low redshifts. At z > 0.6, the most massive and compact systems (R$_{\rm e}^{\rm maj}$ < 2 kpc) present low values of the spin parameter. In contrast, at lower redshift, this feature disappears. If we compare our sample to results on the local Universe, we see that our galaxies at z < 0.6 resemble more the population at z $\sim$ 0, compared to our galaxies at z > 0.6, in terms of mass and sizes \citep{2018MNRAS.477.4711G}. Nevertheless, our sample lacks large (R$_{\rm e}^{\rm maj}$ > 10 kpc) and massive galaxies, as seen in the MaNGA studies \citep{2018MNRAS.477.4711G}, where SRs densely populate the mass-size diagram in the local Universe. 

    \begin{figure}
    \centering
       \includegraphics[width=9.8cm]{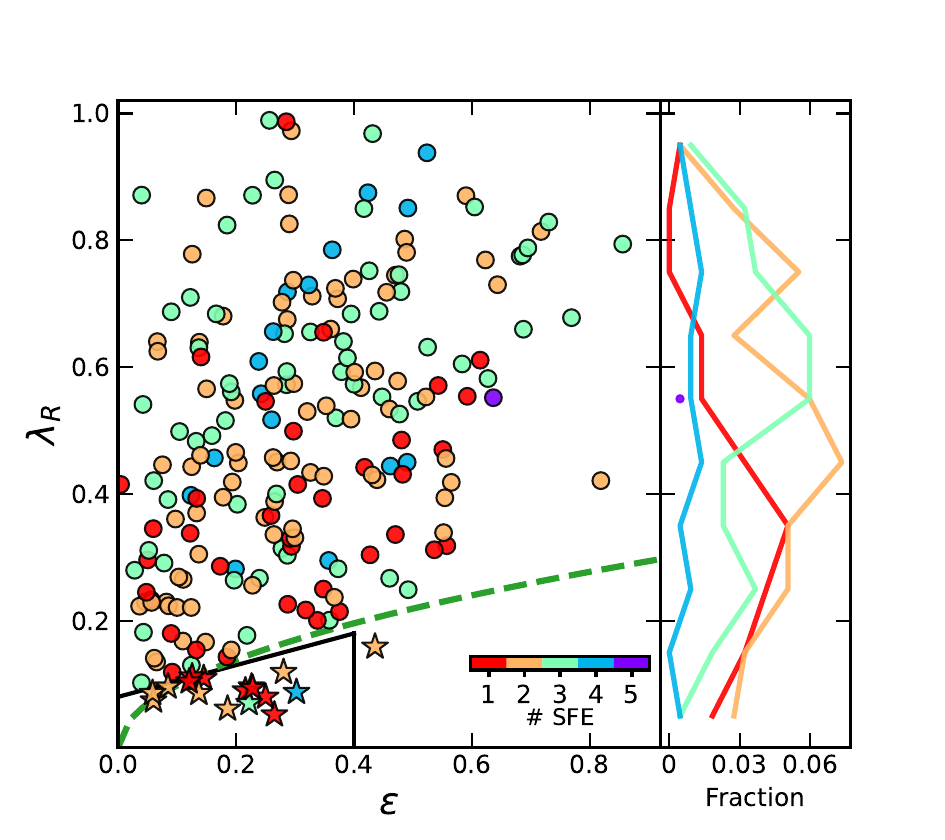}
         \caption{$\lambda_{R}$ as a function of the galaxy's ellipticity coloured by the number of star-forming episodes. The star symbols corresponds to the slow rotators in the sample. The right panel shows the fraction of galaxies as function of the spin parameter with different number of SFEs (same colours as on the legend in the left panel).}
         \label{fig:lam_sfe_paper3}
    \end{figure}
\subsection{Star-forming episodes and spin parameter}\label{sec:pop_kin_paper3}

\begin{figure}
    \centering
       \includegraphics[width=7.5cm]{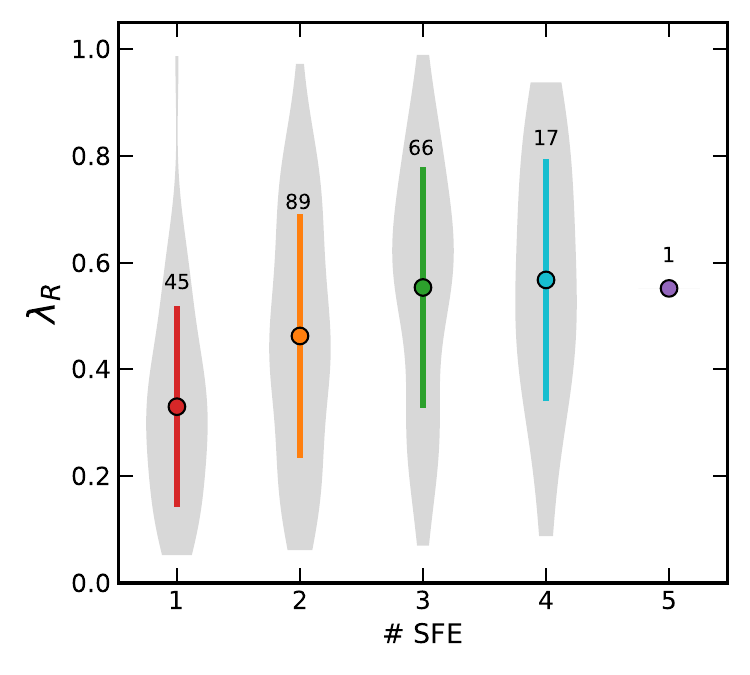}
         \caption{Spin parameter as a function of the number of star-forming episodes. The violins show the $\lambda_{R}$ distribution with the number of galaxies displayed at the top of each set. Circles correspond to the mean of each distribution and colour bars to the 1$\sigma$ distribution.}
         \label{fig:sfe_lam_paper3}
    \end{figure}

In \cite{2025A&A...702A.146M}, we derived the galaxy's SFR as a function of cosmic time from the spectral fitting. Using a combination of exponential and Gaussian functions, we model the changes in the shape of the SFR as a function of time. Thus, using this method, we identified the main star formation events. Figure \ref{fig:lam_sfe_paper3} shows the spin parameter and galaxy's ellipticity coloured by the number of star-forming episodes (SFEs) that galaxies experience throughout their lifetime. We notice that galaxies with low spin parameter assemble their stellar mass in fewer events of star formation, with SFEs $\leq$ 2. Moreover, we also notice that almost all slow rotators (with the exception of two galaxies) assemble their mass in SFE $\leq$ 2, as well . This might not be surprising, as galaxies with one or two main events of star formation dominate the sample as well. However, systems with one SFE are more confined to lower values of spin parameter, $\lambda_{R}$ < 0.4. On the contrary, galaxies with more than one SFE dominate the sample at $\lambda_{R}$ > 0.4. This moderate tendency is better appreciated in Fig. \ref{fig:sfe_lam_paper3}, which shows the distribution of galaxies with different spin parameters as a function of the number of SFEs. Galaxies with two and three SFEs are predominant in our sample, representing in total $\sim$71$\%$ of our galaxy's population. We observed that the spin parameter's distributions are centered toward higher values of $\lambda_{R}$ when the number of SFEs increases. Nevertheless, due to the large scatter, this relation between the number of SF events and the spin parameter is at most tentative. 

The fact that multiple SFEs are more likely to occur in FRs, and that more than 85$\%$ of galaxies experience two or more SFEs, should be connected to the results of \cite{2019A&A...631A..87V}, who found that at z $\sim$ 1, only about 20$\%$ galaxies experience mergers. This means that a large fraction of SFE are fueled by in-situ gas or pure gas accretion (not galaxy mergers, \citealp{2025A&A...702A.146M}). Furthermore, assuming that the tentative difference in the number of SFE between SRs and FRs can be taken at face value, we can conjecture that SFEs connected to gas accretion are more likely to form (or maintain) a FR. Moreover, the formation of SRs is connected to processes that do not involve SF, as is also seen at low-z and in simulations (e.g., \citealp{2014MNRAS.444.3357N,2026A&A...708A..15K}), suggesting that the multiple SFEs are not related to the formation of SRs. However,  simulations of high-z (z > 2.5) galaxies \citep{2025ApJ...979...15K} produce SRs that are both quiescent and star-forming, indicating that the processes of forming SRs might be different at earlier epochs. Nevertheless, most of the additional SFEs in our sample occus at z < 2.5 and might not be relevant to the processes described in these simulations.

\begin{figure}
    \centering
       \includegraphics[width=9.cm]{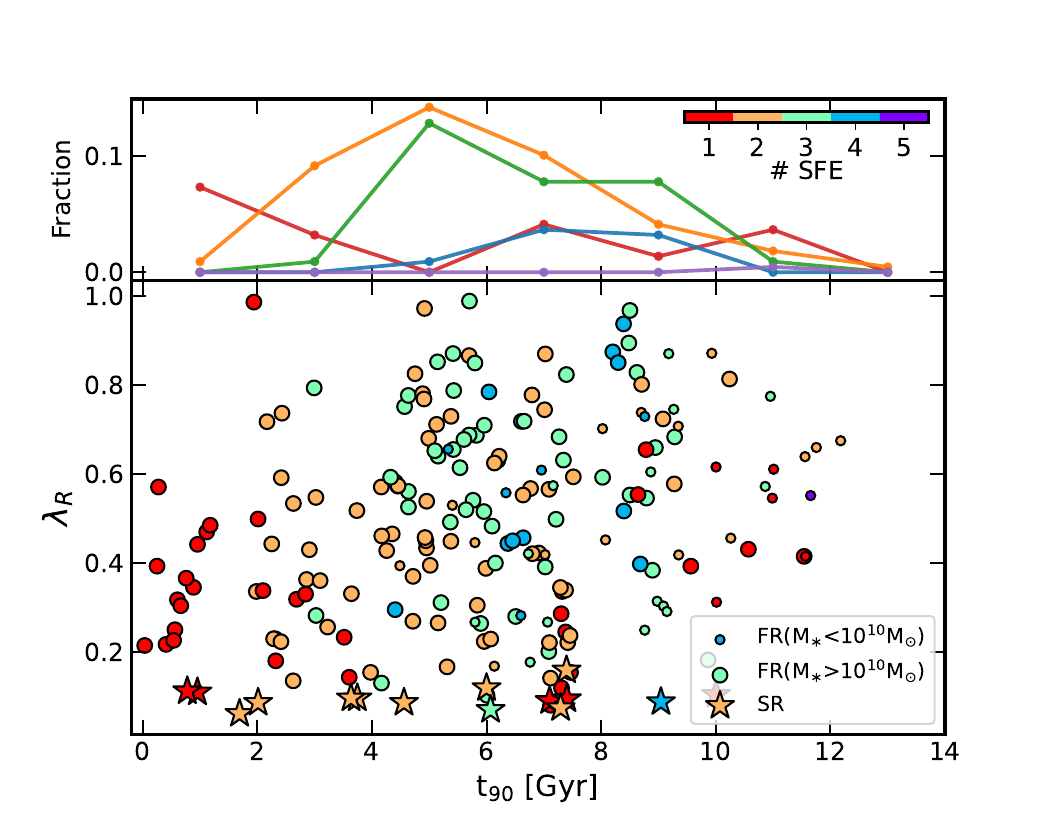}
         \caption{Spin parameter as a function of the formation timescale $t_{90}$. Colours correspond to the number of SFEs. The top panel shows the fraction of galaxies with various number of SFE as a function of $t_{90}$. Stars correspond to slow rotators and circles to fast rotators divided in two mass bins, described in the legend of the figure.}
         \label{fig:t90_lam_sfe_paper3}
    \end{figure}

In order to study the existence of a relation between the timescale formation of galaxies and stellar angular momentum in our sample, in Fig. \ref{fig:t90_lam_sfe_paper3} we present the spin parameter as a function of t$_{90}$, coloured-coded by the number of SF events. We observe that the formation timescale reaches a peak at around 6 Gyr. Most of the sample galaxies had assembled their masses when the Universe was $\sim$6 Gyr old or around z = 1, and at that redshift, only half of the SRs in our sample had finished their star-forming mass assembly. This result is not surprising, given that our sample is dominated by galaxies at z $\sim$ 0.7. Additionally, we are biased towards more massive systems at higher redshifts, and more massive galaxies assemble their mass faster, as showed in \cite{2005ApJ...621..673T,2010MNRAS.404.1775T, 2015MNRAS.448.3484M,2025A&A...702A.146M}. Moreover, we also see that low-mass FRs (log$_{10}$(M$_{\star}$/M$_{\odot}$) < 10), shown by smaller circles in Fig. \ref{fig:t90_lam_sfe_paper3}, form their mass in longer timescales, t$_{90}$ > 6 Gyr, and present mostly more than one episode of star formation. 

Figure \ref{fig:t90_lam_sfe_paper3} shows that all SRs have assembled their mass by z $\sim$ 0.3 (or when the Universe is 10 Gyr old), and are not star-forming (except the two low-mass, low-z SRs). It is not clear whether the transition of a galaxies from FR to SR is directly related to quenching. This scenario could be true, but there is evidence that SRs can have low levels of star formation (for example: NGC1222 in \cite{2020A&A...635A.129K} ; MaNGA sample from \cite{2025NatAs...9..165W}; simulations from \cite{2025ApJ...979...15K}; the star-forming SR in our sample). However, these are special cases, in the sense that these galaxies might just be going through a phase of low $\lambda_{R}$ and it is not clear what they will be after the star formation is completed. This means that typical SRs are galaxies that finished assembling their mass via star formation, some time before becoming SRs.

\subsection{Environmental indicators}\label{sec:env_paper3}

We quantify the environment in our sample using the global density estimator $\eta$ \citep{2013MNRAS.436L..40N}. We follow a similar approach as \cite{2024A&A...683A.205E}, where a friends-of-friends (FoF) algorithm was used to identify structures and galaxy groups. The parameter $\eta$ accounts for the level to which galaxies are bound to their respective groups and is defined as: 
    \\
    \begin{equation}
        \eta = \frac{\mid \Delta v \mid}{\sigma_{g}} \frac{\Delta r}{R_{200}}
    ,\end{equation} 
    \\

where $\Delta v$, $\Delta r$, and $\sigma_{g}$ are the velocity of the galaxy within the group, the projected distance to the group centre, and the velocity dispersion of group members, respectively. R$_{200}$ is the radius at which the group density is equal to 200 times the Universe critical density, and it holds the assumption that the group is virialized. Thus, $\eta$ is a measurement for how dynamically tied a galaxy is to its group. Galaxies with $\eta$ < 0.1 are considered early accreted by their group. If they already passed the pericentre of their orbit once 0.1 < $\eta$ < 0.4 is expected, and if they have recently been accreted by the group, 0.4 < $\eta$ < 2 is anticipated \citep{2013ApJ...768..118N}. Our sample also contains systems that are not part of any galactic structure or group, thus isolated systems, namely field galaxies.

As mentioned in Sect. \ref{sec:pop_kin_paper3}, we divided the sample into two redshift bins. We considered field systems, galaxies that were not placed in any group by the FoF algorithm, and galaxies identified in a group, but with members N $\leq$ 3 \citep{2023A&A...677A.143M}. Of the galaxies at low redshift (z < 0.6), 88 systems are in group structures, which represents $\sim$72$\%$ of the galaxies in their redshift bin, and $\sim$40$\%$ of the overall galaxy sample. On the other hand, in the higher redshift bin, almost all systems are in groups, with only 7 isolated galaxies. Approximately 82$\%$ (14/17) of our SRs are in rich groups ($N_{members}$ > 10), two SRs are field systems and one is in a group with $N_{members}$ = 4.

\begin{figure}
    \centering
       \includegraphics[width=9.cm]{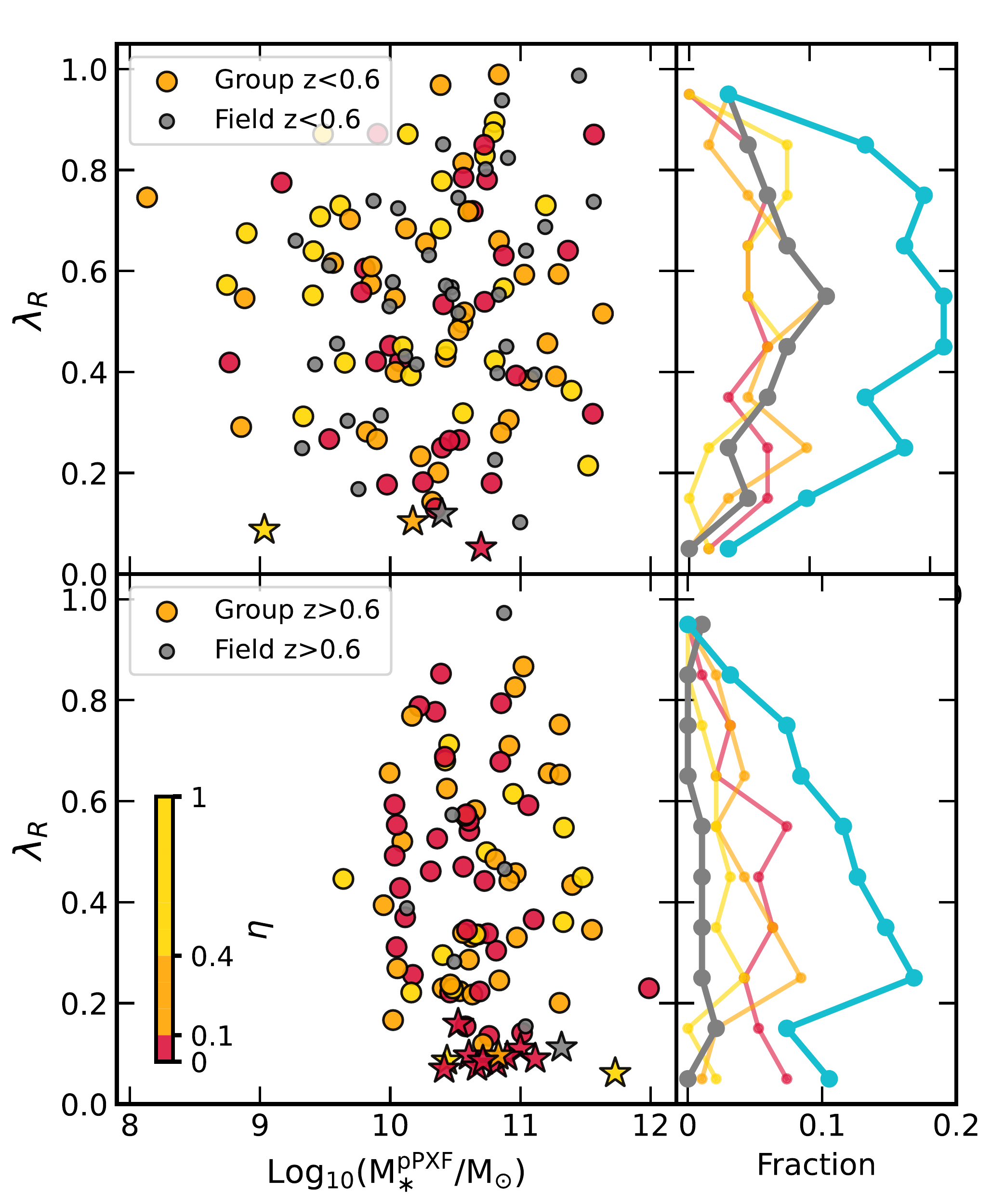}
         \caption{Spin parameter as a function of the stellar mass, colour-coded by the global environmental indicator $\eta$. Stars and circles show slow and fast rotators, respectively. Grey small circles and stars show isolated systems or field galaxies. Top and bottom panels show galaxies with redshifts z < 0.6 and z > 0.6, respectively. Right panels show the corresponding fraction of galaxies as function of $\lambda_{R}$ following the same colour scheme as of the left panels. Sky blue lines correspond to the sum of all group galaxies, thus the addition of crimson, orange and yellow circles and stars.}
         \label{fig:mass_lam_eta_paper3}
    \end{figure}

In Fig. \ref{fig:mass_lam_eta_paper3}, we present the spin parameter as a function of the stellar mass, colour-coded by $\eta$ for the two redshift bins, with their corresponding fraction of galaxies as a function of $\lambda_{R}$. 
In the right panel, we observe that in both redshift bins, the sample is dominated by galaxies in groups, shown in the figure by the blue continuous line. At z < 0.6, the fraction of group galaxies peaks at higher spin parameter $\lambda_{R}$ > 0.4, while at z > 0.6, a high fraction of the galaxies present $\lambda_{R}$ $\sim$ 0.3.

Moreover, we analyse the kinematic morphology-density relation \citep{2011MNRAS.416.1680C}. In our sample, the higher-redshift bin is dominated by massive galaxies, which reside in more dense environments (lower $\eta$) compared to group galaxies at z < 0.6. We notice that low-$\lambda_{R}$ galaxies prefer high-density (group) regions. A similar trend is observed in the z $\sim$ 0 sample of \cite{2021MNRAS.508.2307V} (also in \citealp{2019191005139G}). This study used $\sim$1800 ETGs and LTGs with stellar masses of log$_{10}$(M$_{\star}$/M$_{\odot}$) > 9.5, finding that galaxies in high-density environments present lower values of the spin parameter. They also observed that galaxies with $\lambda_{R}$ < 0.2, are mostly group, cluster centrals, and their satellite systems. Our results at z < 0.6 are comparable to \cite{2021MNRAS.508.2307V}; we find that the fraction of group galaxies, with $\lambda_{R}$ < 0.2, is higher than field systems.

The impact of the environment on the stellar kinematics properties of massive quiescent galaxies at higher redshifts has been studied using data from the LEGA-C survey. \cite{2020ApJ...890L..25C} analysed the relation between the (1+$\delta$) overdensity and the rotational support (V/$\sigma$) of galaxies at z $\sim$ 0.8, finding similar results as galaxies in the nearby Universe. This is consistent with previous results of galaxies in the local Universe; the rotational support of massive galaxies depends mainly on the stellar mass; more massive galaxies are found in denser environments. Notwithstanding the differences in the characterisation of the environments in our sample and that of \cite{2020ApJ...890L..25C}, we observe a qualitatively similar trend for our high-z galaxies (z > 0.6). The LEGA-C results \citep{2020ApJ...890L..25C}, based on V/$\sigma$, are similar to ours in the sense that more massive galaxies have lower rotation support. This is also expected from the z = 0 studies, but we see an environmental dependence as well. In the bottom panel of Fig. \ref{fig:mass_lam_eta_paper3}, we notice that more massive systems, which are in the densest (group) environments (red circles), have a wide range of stellar angular momentum. However, in dense environments, the frequency of SRs (star symbols) is higher than in other environments, given that our galaxy sample is biased towards groups at z $\sim$ 0.7 (by construction of the MAGIC survey).

\subsection{Slow rotators}\label{sec:slow_rotators_paper3}

Our sample contains 17 slow rotators, representing $\sim$8$\%$ of the total number of galaxies. These slow rotators range from redshift z $\sim$ 0.3 to z $\sim$ 0.94. Three of them were already identified in \cite{2024A&A...688A..75M,2026A&A...707C...4M}, thus 14 are SRs reported in this work, and 13 of them are galaxies within the MAGIC survey footprint. As the MAGIC survey aimed to study massive galaxies' structures, it is not surprising the presence of a higher number of SRs in denser environments.

\begin{figure}
    \centering
       \includegraphics[width=9.cm]{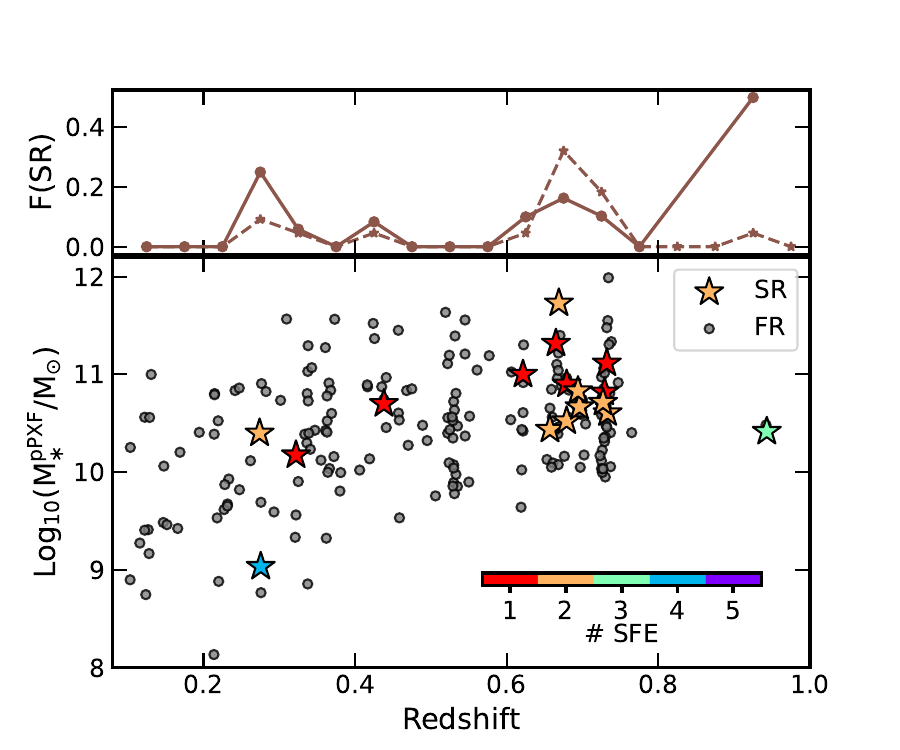}
         \caption{Stellar mass as a function of the redshift for the sample. Circles show fast-rotating galaxies and stars slow rotators coloured-coded by the number of star-forming events. The top panel show the fraction of slow rotators as a function of the redshift; the solid line corresponds to the relative fraction of SR according to the number of galaxies in the redshift bin, while the dashed line the absolute fraction of SR, when compared to all galaxies in the sample (the absolute fraction is amplified by a factor of 10 for better visualization).}
         \label{fig:z_mass_sfe_paper3}
    \end{figure}

In Fig. \ref{fig:z_mass_sfe_paper3}, we present the redshift distribution of SRs in bins of 0.1 redshift, plotted against their stellar mass. We observe that the majority of SRs are systems around z $\sim$ 0.7, representing almost 20$\%$ of the galaxies in that redshift bin. Most of the SRs (16/17) are massive galaxies with stellar mass M$_{\star}$ > 10$^{10}$M$_{\odot}$, formed in either one or two SFEs. Therefore, the SRs in our sample are mostly dominated by a few distinct stellar populations. Our sample contains one low-mass SR, Galaxy 72. The system is at z $\sim$ 0.28, has a stellar of M$_{\star} \sim$ 10$^{9}$ M$_{\odot}$ and maintains some level of star-formation (see Fig. \ref{fig:mass_sfr_lam_paper3}). 

The mass-size diagram compared to local surveys is presented in the top panel of Fig. \ref{fig:mass_size_arrow_paper3}. Galaxy 72 has a size of R$_{\rm e}^{\rm maj} \sim$ 3 kpc (see Fig. \ref{fig:mass_size_arrow_paper3}) and has four detectable stellar populations. 

\cite{2020ApJ...890L..25C} found that the fraction of low V/$\sigma$ galaxies at lower masses (log$_{10}$(M$_{\star}$/M$_{\odot}$) < 11.25) is similar for LEGA-C galaxies and local samples (ATLAS$^{\rm 3D}$ and MASSIVE); the fraction of SRs is driven by the mass, where at higher masses the fraction of slow rotators is larger, independently of the environment. However, at z $\sim$ 0.8, the fraction of SRs with mass log$_{10}$(M$_{\star}$/M$_{\odot}$) > 11.25 follows a trend with the environment, being the most dense environment (1+$\delta$ $\geq$ 3.5) more favorable for finding SR ($\sim$90$\%$).

In the case of our slow-rotating population, we do not observe a clear trend with the environment. However, we observed that the majority of SRs (15/17) reside in group environments at z > 0.6. We also find a similar trend of increasing fraction of SRs at higher masses (see the bottom panel of Fig. \ref{fig:mass_size_arrow_paper3}), as it is found for galaxies in the nearby Universe \citep{2021MNRAS.508.2307V,2018MNRAS.477.4711G}.

\subsubsection{What are the mechanisms for the formation of slow rotators?} Simulations like EAGLE and HYDRANGEA at z = 0 \citep{2018MNRAS.476.4327L}, reported that $\sim$70$\%$ of SRs in the local Universe have experienced at least one merger with mass ratio $\geq$ 0.1. While dry mergers are twice as common as wet mergers, they decrease the value of $\lambda_{R}$, in contrast to wet ones that increase the stellar angular momentum. \cite{2022MNRAS.509.4372L}, observed that SRs at z = 0 that have not experienced major or minor merging events are recently quenched, more compact (r$_{50, \rm r-band}$< 4 kpc), and contain specific characteristics in their stellar populations.

At higher redshift, \cite{2018MNRAS.480.4636S}, using the cosmological simulation Magneticum Pathfinder, showed that the SR population starts appearing in a statistically important fraction around z $\sim$ 2. \cite{2025ApJ...979...15K}, analysed the stellar kinematics properties of star-forming and quiescent galaxies between z $\sim$ 5 and z $\sim$ 3, using the Magneticum Pathfinder simulation as well. They found no relation between the star-forming state of galaxies and their stellar kinematics. Their slow-rotating population emerges between z = 4 and z = 2, and their quenching process is not connected to a decrease in rotational velocity or an increase in the velocity dispersion. Surprisingly, from the point of view of properties of SRs at z = 0, most of their SRs are star-forming. At z >3 , when the Universe is about 2 Gyr old, it is not trivial that there is sufficient time for dry mergers to take place. Therefore, the formation of SRs at those early epochs likely do not follow the same scenarios as at z = 0. While it is possible to form a slow rotator from a merger of gas-rich galaxies \citep{2010MNRAS.406.2405B,2011MNRAS.416.1654B}, it requires a special merger configuration, and it is not likely to happen often.

\begin{figure}
    \centering
       \includegraphics[width=8.5cm]{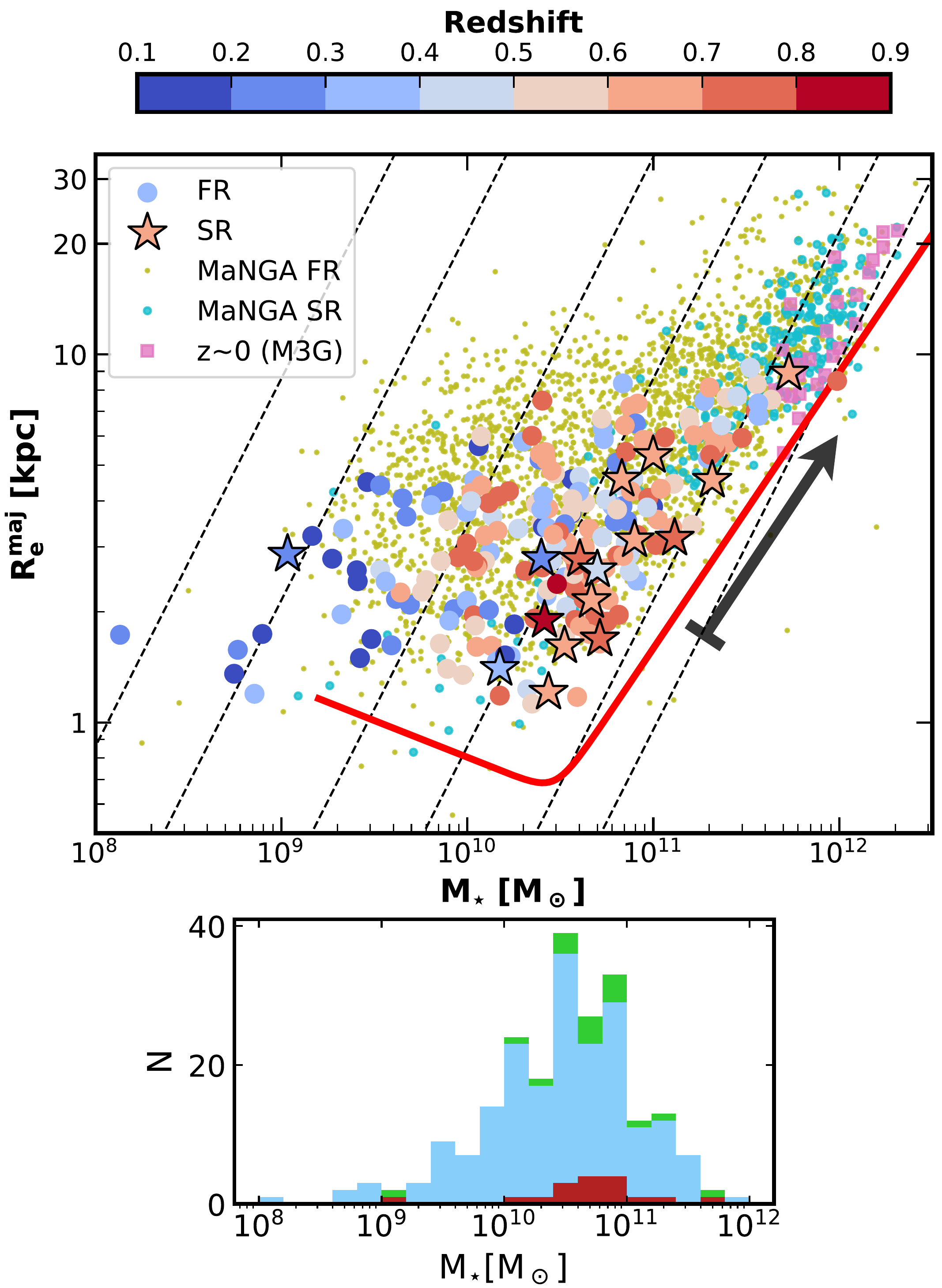}
         \caption{\textit{Top:} plane of effective radius (R$^{\rm maj}_{\rm e}$) versus stellar mass coloured-coded by the redshift for the 218 galaxies. Stars and circles show slow and fast rotators in the sample, respectively. Olive and cyan dots show fast and slow rotators at z $\sim$ 0, respectively, from the MaNGA survey, presented in \cite{2018MNRAS.477.4711G}. Pink squares correspond to galaxies from the MUSE Most Massive Galaxies (M3G) in the nearby Universe \citep{2026A&A...708A..15K}. Dashed-black lines show constant velocity dispersion of 10, 20, 50, 100, 200, and 300 km s$^{-1}$ from left to right, derived by the virial mass estimator M = 5R$^{\rm maj}_{\rm e}$ $\sigma^{2}$/G. The solid red line indicates the zone of exclusion (ZOE) from equation (4) of \cite{2013MNRAS.432.1862C}. The black arrow indicates approximate galaxy growth via gas-poor (dry) merging. \textit{Bottom:} stellar mass distributions of all galaxies in the sample, fast rotators, and slow rotators, in green, light blue, and red, respectively. }
         \label{fig:mass_size_arrow_paper3}
    \end{figure}

Figure \ref{fig:mass_sfr_lam_paper3} shows the star formation rate as a function of the stellar mass, where the SRs are highlighted by star symbols. The SFR in this figure comes from SED fitting computation \citep{2014ApJS..214...24S,2024A&A...683A.205E}, in consequence, it is tracing the most recent star formation in the galaxies, around the last $\sim$100 Myr. Most of our slow-rotating galaxies are quenched systems, with the exception of 3 systems (Galaxies 72, 168, and 215), which are actively forming stars. We observe in Fig. \ref{fig:mass_size_arrow_paper3} that our slow-rotating population spans a wide range of sizes, from 1 to 10 kpc. Thus, our SRs are SRs at least from the last SFE, but it is possible that they experienced dry merging in the mean time. According to the characterization of our slow-rotating population, given their amount of SF events, current SFR, stellar masses, sizes, and stellar population parameters, their most likely formation scenario is through mergers that destroy their disk and regular kinematics, which did not trigger important episodes of star formation.

As the MUSE surveys used in this study are flux-limited, our sample is biased toward the largest and most massive galaxies at higher redshift. 
Hence, the most massive population of galaxies observed at z $\sim$ 0 is not represented in our sample because these systems have not yet formed. The inter-z galaxies in our sample are smaller and less massive than nearby galaxies, because they are still growing. Galaxies in our sample should grow between about 2-20 times bigger to become the most massive galaxies in the nearby Universe (see M3G galaxies in Fig. \ref{fig:mass_size_arrow_paper3}). Regardless, they will either passively evolve, keeping their position in the mass-size diagram, or grow diagonally through dry mergers (black arrow in Fig. \ref{fig:mass_size_arrow_paper3}). Such evolution (along the black line) is expected from dry mergers, as they will increase the mass and size linearly, but keep the velocity dispersion nearly constant (e.g. \citealp{2009ApJ...699L.178N,2009ApJ...697.1290B}). Wet mergers could be possible, but do not seem to be likely because they already seem to be pilling towards the zone of exclusion (ZOE).

Some models suggest that z $\sim$ 0 massive galaxies are expected to experience a few (1-2) major dry mergers (e.g \citealp{2007MNRAS.375....2D}). Observationally, this is supported by massive elliptical galaxies having low-surface-brightness cores, which is related to their supermassive black hole (SMBH) mass \citep{2009ApJS..182..216K}. During galaxy mergers, BH binaries form naturally and are believed to form cores \citep{1980Natur.287..307B,2001ApJ...563...34M}. The BH binaries expel stars from the central region of the galaxy, resulting in a lack of stars in the volume near the BHs. This process results in a decrease in the central surface brightness of the galaxy \citep{2012ApJ...744...74G,2018ApJ...864..113R}. In addition, the mass of the resulting SMBH scales linearly with the number of dry mergers. By knowing how much mass is missing from the cores of massive ellipticals, the number of dry mergers can be estimated \citep{2006ApJ...648..976M}. Some estimates indicate that the typical number of mergers for the formation of these cores ranges between $N \sim$ 2-4, but even $N\sim$ 6-10 is possible \citep{2013MNRAS.429.2924H,2019ApJ...886...80D}. \cite{2012MNRAS.425.3119H} studied the impact of collisionless major and minor mergers, in the context of compact massive early-type galaxies at z $\sim$ 2. They found that minor dry merging (mass ratio 1:10) is a more efficient mechanism in the production of massive elliptical systems, significantly affecting their size growth. \cite{2024MNRAS.535.1202R} showed the importance of the supermassive black hole (SMBH) binaries in expanding the size of galaxies, especially in minor merger cases, where the most effective way to increase the size is when a sequence of minor mergers (with SMBHs) follows a single major merger. Overall, these numbers are in agreement with the estimated size and mass evolution for the SRs in our sample.

Therefore, the inter-z slow-rotating population in our sample will continue evolving in mass and size and start to build up the most massive and largest population of galaxies at z $\sim$ 0. Consequently, we conclude that the slow rotators in our sample are consistent with being the progenitors of the most massive slow rotators observed at z $\sim$ 0, which later grow through dry mergers.

\subsubsection{Slow rotation fraction}

\begin{figure}
    \centering
       \includegraphics[width=9.0cm]{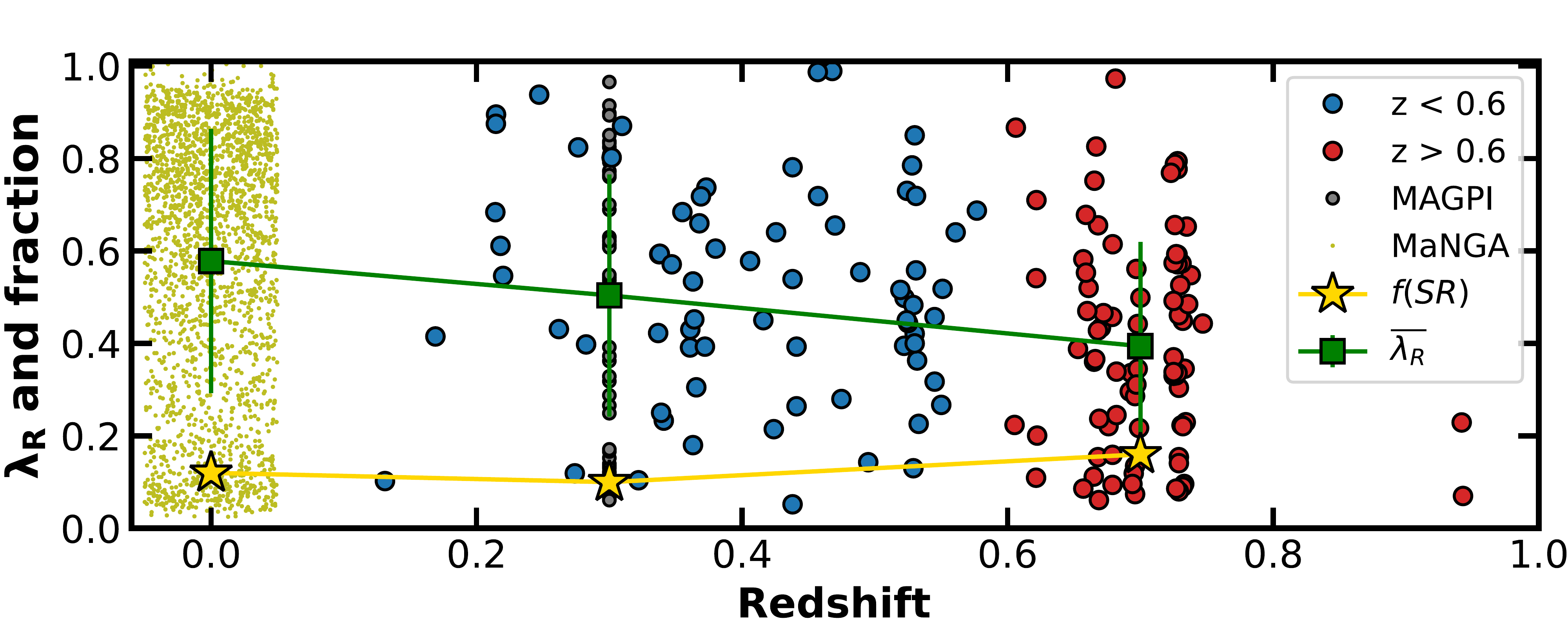}
         \caption{Spin parameter as a function of the redshift for galaxies with stellar mass M$_{\star} \geq$ 10$^{10}$ M$_{\odot}$. Blue and red dots correspond to galaxies of our sample at z < 0.6, and z $\geq$ 0.6, respectively. Olive dots show galaxies at z $\sim$ 0 from the MaNGA survey, presented in \cite{2018MNRAS.477.4711G}. Grey circles are galaxies from the MAGPI survey at z $\sim$ 0.3 \citep{2024MNRAS.531.4602D}. Golden star symbols represent the fraction of SRs at z $\sim$ 0 (MaNGA), at z $\sim$ 0.3 (MAGPI and our galaxies at z < 0.6), and at z $\sim$ 0.7 (our galaxies at z > 0.6). Green squares correspond to the mean spin parameter for the redshift bins mentioned previously. }
         \label{fig:frac_sr}
    \end{figure}

We analyse the fraction of slow rotators at different redshifts between 0 < z $\lesssim$ 1. We select galaxies with stellar mass M$_{\star} \geq$ 10$^{10}$ M$_{\odot}$ from different surveys: at z $\sim$ 0, galaxies from the MaNGA survey \cite{2018MNRAS.477.4711G}; at z $\sim$ 0.3 galaxies from the MAGPI survey \citep{2024MNRAS.531.4602D} and systems in our sample with z < 0.6, and at higher redshift galaxies from our sample with z > 0.6. These samples are shown in Fig. \ref{fig:frac_sr} with olive, grey, blue, and red circles, respectively. The fraction of SRs at different redshifts is shown as golden stars. At z $\sim$ 0, 12$\%$ (273/2289) of the sample corresponds to SRs. At z $\sim$ 0.3, 9/51 MAGPI galaxies are slow-rotating, and 3/71 of our galaxies at z < 0.6 are SRs, which, when adding both samples, corresponds to a fraction of SRs of $\sim$10$\%$. At z > 0.6, the fraction of SRs is 16$\%$ (13/82). Therefore, we do not observe an evolution in the fraction of slow rotators from redshift z = 1 to z = 0.

The green squares in Fig. \ref{fig:frac_sr} show the mean spin parameter as a function of the three redshift bins (z $\sim$ 0, z $\sim$0.3, and z > 0.6). Even though there is no noticeable trend in the fraction of SRs, we observe a mild tendency for the sample in the local Universe to present, on average, higher values of $\lambda_{R}$. This tendency is a consequence of the inclusion of lower mass galaxies (M$_{\star} \leq$ 10$^{10}$ M$_{\star}$) in surveys at z $\sim$ 0. This result is in contrast to \cite{2026ApJ..1009...87M}, which found a gradual decrease in $\lambda_{R}$ with decreasing redshift. However, the fact that our galaxy sample is biased towards groups at z $\sim$ 0.7 makes the comparison of the fraction of SRs as a function of the redshift biased.

\subsection{Sample biases}\label{sec:bias_paper3}

As mentioned previously, our sample is flux-limited. As our sample spans a wide redshift range, when we move to higher-z, only the brightest objects fulfill the S/N requirements for selecting galaxies (see Sect. \ref{sec:selection_paper3}). Additionally, a second effect is present in the sample. The physical size of galaxies plays an important role in the creation of the kinematic maps, as well as in the implementation of the seeing corrections to the spin parameter. As we move to higher redshifts, galaxies are less massive, smaller and appear smaller on the sky. This limits our sample to include only the most massive, brightest, and largest galaxies at higher redshifts, an effect that we observed in the bottom panel of Fig. \ref{fig:mass_size_paper3}. 

As the MUSE-Wide survey is a blind survey, we also do not target galaxies with specific characteristics at low redshifts. With the integration time of MUSE-Wide (1h), we do not detect a large number of systems with stellar masses M$_{\star}\lesssim$ 10$^{10}$ M$_{\odot}$ at low redshifts. The lack of an abundant number of these galaxies at low-z (due to their brightness and sizes) most likely drives the absence of a peak in the spin parameter distribution (see Fig. \ref{fig:eps_lam_correc_paper3} and Fig. \ref{fig:spin_obs_paper3}) at $\lambda_{\rm R}$ $\sim$ 0.9, as observed in z $\sim$ 0 samples \citep{2018MNRAS.477.4711G}. As most of our sample galaxies have stellar masses 10 $\lesssim$ log$_{10}$(M$_{\star}$/M$_{\odot}$) $\lesssim$ 11 (see Fig. \ref{fig:z_mass_sfe_paper3}), we also lack the peak in the spin parameter distribution at lower values ($\lambda_{\rm R}$ $\sim$ 0.1), where a high fraction of massive systems populate the $\epsilon$ - $\lambda_{\rm R}$ diagram in \cite{2018MNRAS.477.4711G}. The combined properties of our sample, at both high and low redshifts, and their overall limited number of galaxies (218), result in the non-presence of a bimodal distribution in the stellar angular momentum.

Moreover, it is not surprising that the fraction of SRs found in the MAGIC survey (13/17) is much greater than the one including all the surveys within the MUSE-Wide footprint at high-z, given that the MAGIC survey offers similar environmental conditions where most of the slow-rotating population is found in the nearby Universe. As similarly as the MAGPI survey is targeting massive central galaxies at z $\sim$ 0.3. 

\section{Conclusions}\label{sec:conclusions_paper3}

We measured the stellar angular momentum, parametrized by the spin parameter, for 113 galaxies in the MAGIC survey, and together with galaxies from \cite{2024A&A...688A..75M,2026A&A...707C...4M}, we presented the resolved stellar kinematics for a sample of 218 galaxies at redshifts 0.1 $\lesssim$ z $\lesssim$ 1. The sample covers galaxies with stellar masses of M$_{\star}$$\sim$ 10$^{8}$-10$^{12}$ M$_{\odot}$ and from star-forming to quiescent systems.

Our sample \footnote{The table presenting the properties of the galaxy sample is available at CDS.} is dominated by round systems with ($\epsilon \leq$ 0.3) and galaxies with $\lambda_{R}$ < 0.8. The sample contains 17 slow rotators, which corresponds to $\sim$8$\%$ of the total number of galaxies in the sample. We identified the SR at the highest redshift to date, Galaxy 198, at z = 0.943. We classified the kinematic maps of the galaxies according to their kinematical resemblance to a stellar disc. We found that $\sim$66$\%$ of the stellar maps exhibit regular rotation, which is in agreement with the number of FRs in the sample. We do not observe a bimodality in the $\lambda_{R}$-$\epsilon$ diagram as observed in samples in the nearby Universe.

Additionally, we use the stellar population parameters and star formation histories for those galaxies derived in \cite{2025A&A...702A.146M}, and analysed the relation between the history of mass assembly and the stellar kinematics, and their dependence on the galaxy's environment, over $\sim$ 6 Gyr in cosmic time. We observed a high fraction of low stellar angular momentum galaxies that experience fewer episodes of star formation during their lifetime. After separating the sample into two redshift bins, we observed that our galaxies at z < 0.6 resemble the population at z $\sim$ 0, in terms of mass and sizes, more than systems at z > 0.6. At z > 0.6, the most massive and smaller galaxies present low values of angular momentum. This behaviour resonates with the most massive galaxies (M$_{\star}$ > 2 $\times$ 10$^{11}$ M$_{\odot}$) in the LEGA-C survey, which generally show no or little rotation. We observed that the spin parameter's distributions are centered toward higher values of $\lambda_{R}$ when the number of SFEs increases. This is a tentative result, but it is consistent with the fact that SRs are expected to be formed from gas-poor interactions (colissionless mergers), and processes that do not involve accretion of gas; FRs, on the other hand, can be formed or maintained through SF process.

We quantified the environment in our sample using the global density estimator $\eta$, which accounts for galaxies in structures and the level of boundness of the system to its groups. We observed evidence for the kinematic morphology-density relation in our subsample at z > 0.6.

We also analysed the fraction of SRs at different redshifts using various surveys, finding no significant evidence for evolution. However, we find that the mean spin parameter is, on average, lower at higher redshifts.  

Through the analysis of the mass-size diagram, compared to surveys in the nearby Universe, we concluded that our SR population is consistent with being the progenitors of the most massive SRs observed at z $\sim$ 0, which later grow through gas-poor (dry) mergers.

\section*{Data availability}

\begin{itemize}
    \item Table \label{sec:sample_table_paper3} is only available in electronic form at the CDS. 

\item The stellar kinematic maps (see Appendix \ref{sec:kin_maps_appendix_paper3}) used in this work are available in \hyperlink{https://zenodo.org/records/23953347?token=eyJhbGciOiJIUzUxMiJ9.eyJpZCI6IjI0YTUwN2Y4LTViYmQtNDY4ZC1hMDA2LTY0ZTQwOTE1MzBkMiIsImRhdGEiOnt9LCJyYW5kb20iOiI1ZDAxYmQ0ODUzZjQ0ZDU2MTlhYjljMmEwNDEwZThmNyJ9.RVXBjyCPX8Ey5iJrvbN3JBb4Hgsm5V9bo0zEdx-YbRIhEiSNIlJ4Gg6apvr4DnP1dqJMpJc_u4yU0rikJsBLfQ}{Zenodo.} 
\end{itemize}

\begin{acknowledgements}
      This work was supported by the German
      \emph{Deut\-sche For\-schungs\-ge\-mein\-schaft, DFG\/} project
      number 4548/4-1.
\end{acknowledgements}

\clearpage

\begin{appendix}

\onecolumn

\begin{figure}[hbt!]
\section{Stellar kinematic maps}\label{sec:kin_maps_appendix_paper3}
\centering
   \includegraphics[width=16cm]{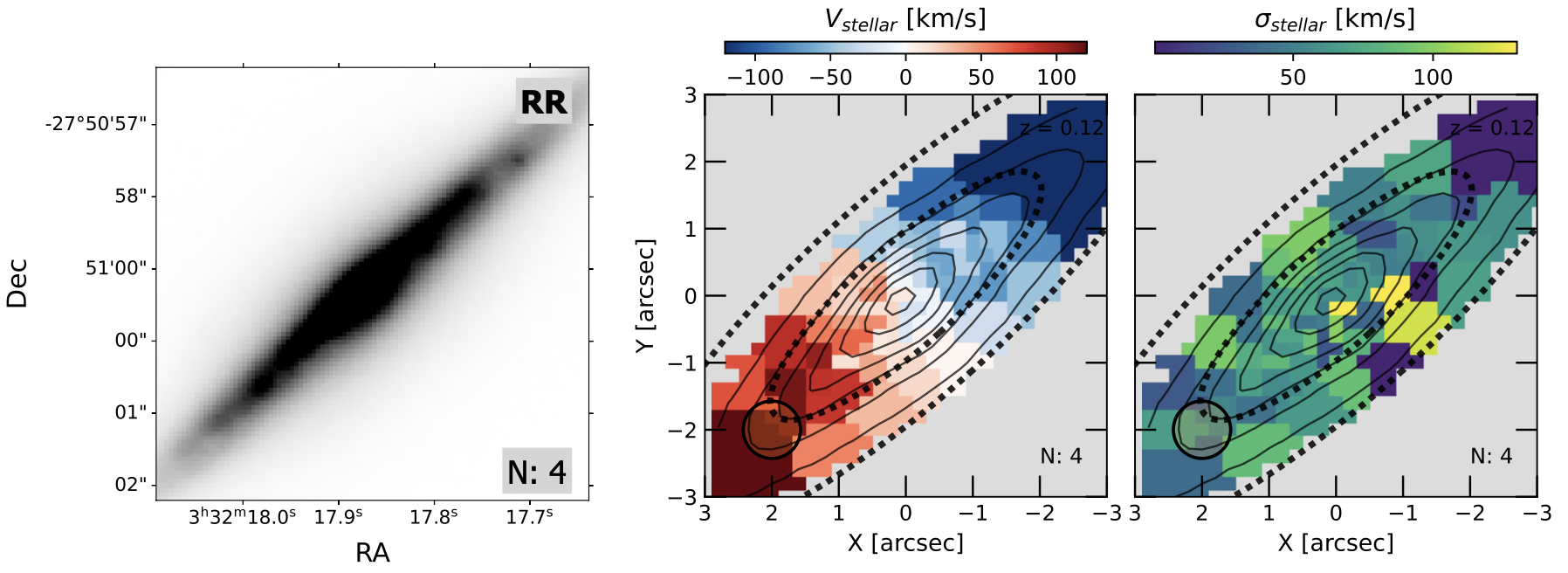}
      \caption{HST image (left), resolved stellar line-of-sight velocity field (middle) and velocity dispersion map (right) of each galaxy in the sample. Kinematic maps are Voronoi binned. The contours are isophotes of the surface brightness from the MUSE white light image plotted with black continuous lines. The dashed thick black lines show ellipses computed using 1Re and 2Re, respectively. The number N in the lower right part of each panel denotes the galaxy index number. Velocity and velocity dispersion ranges are colour-coded (see colour bar at the top of each panel). The orange circles illustrate the size of the PSF (FWHM) in arcsec. The kinematic classification is included in the top right part of the HST image panel. North is to the top and east is to the left. The kinematic maps are available as supplementary material.}
     \label{fig:map_example_paper3}
\end{figure} 

\hspace{1cm}
    \begin{figure}[hbt!]
    \section{Observed spin parameter}\label{sec:spin_obs_paper3}
    \centering
       \includegraphics[width=9.9cm]{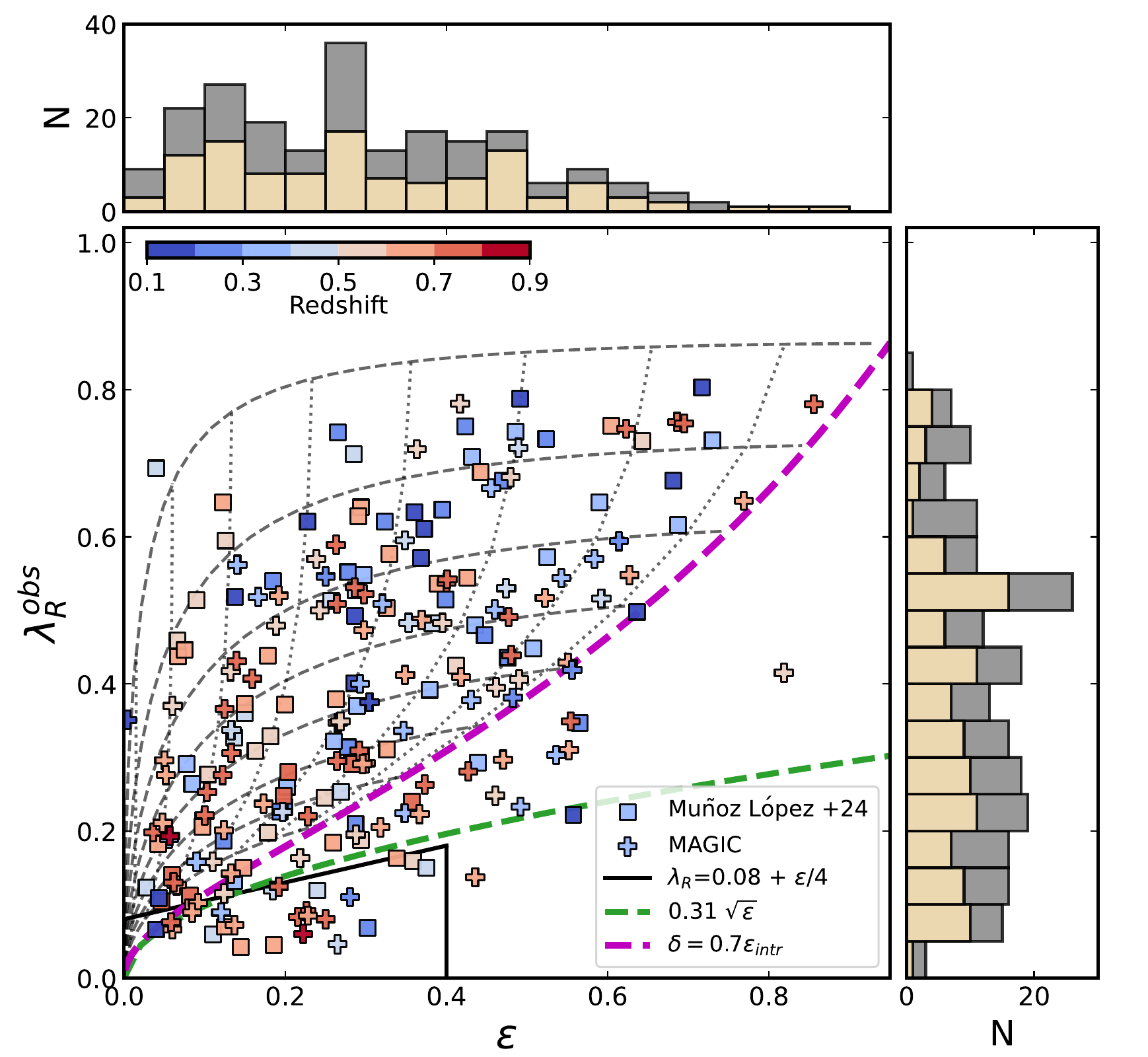}
         \caption{Observed $\lambda_{R}$ as a function of the galaxy's ellipticity coloured by the redshift for the 218 galaxies in the sample. Squares are galaxies presented in \cite{2024A&A...688A..75M}. Crosses correspond galaxies within the MAGIC footprint, presented in this work. The galaxy's ellipticity and spin parameter distributions are shown by the histograms. Grey bars show the sample galaxies from \cite{2024A&A...688A..75M}. Wheat-coloured bars show galaxies from the MAGIC survey \citep{2024A&A...683A.205E}. Full description in Fig. \ref{fig:eps_lam_correc_paper3}}
         \label{fig:spin_obs_paper3}
    \end{figure}
\twocolumn
    \begin{figure*}
    \section{Spin parameter LOESS smoothed}
    \sidecaption
       \includegraphics[width=12.5cm]{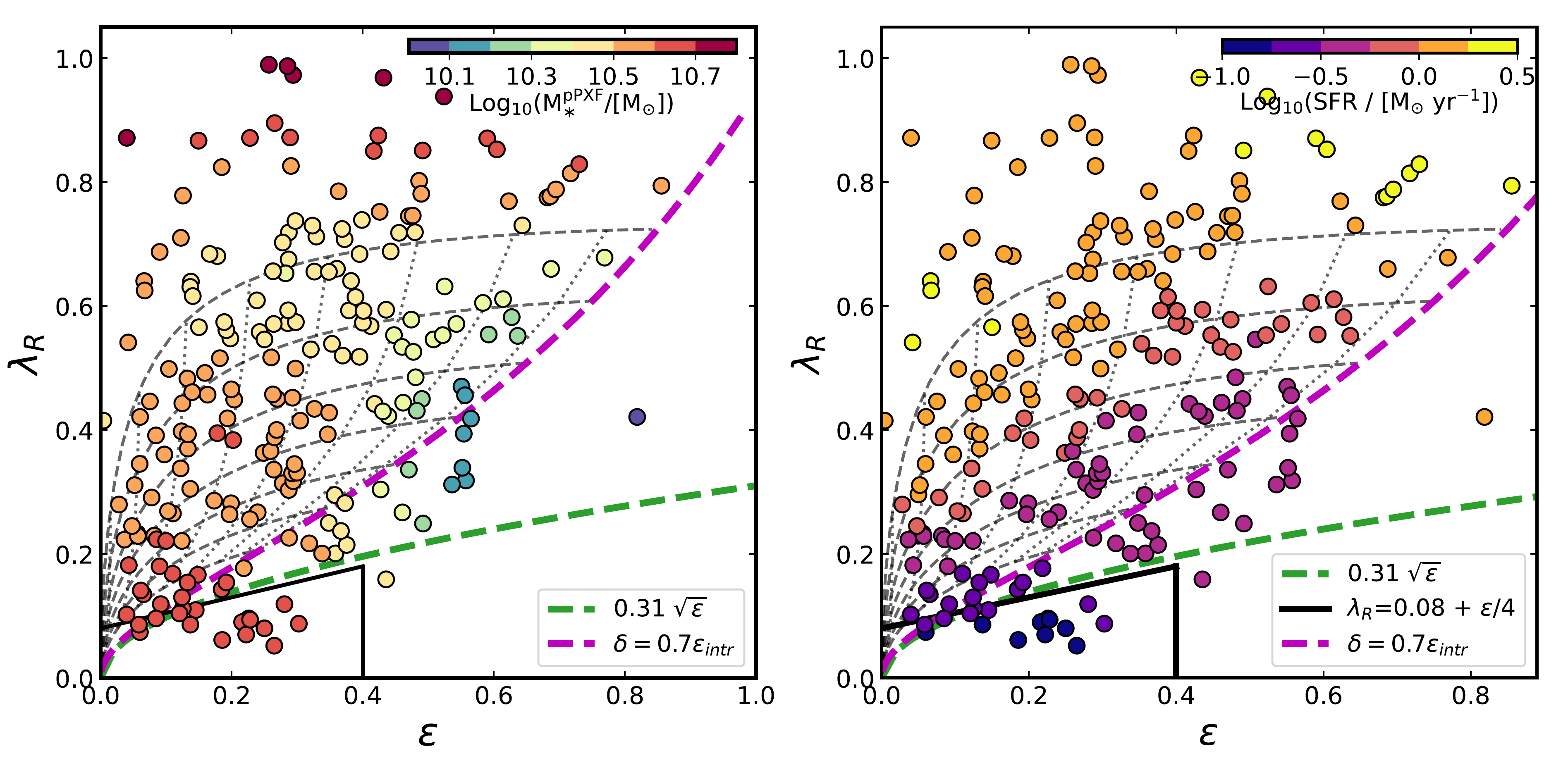}
         \caption{\textit{Left:} $\lambda_{R}$ as a function of the galaxy's ellipticity coloured by the stellar mass. \textit{Right:} $\lambda_{R}$ as a function of the galaxy's ellipticity coloured by the star formation rate. Black, green, magenta, dotted and dashed lines are described in Fig. \ref{fig:eps_lam_correc_paper3}. The stellar mass and the SFR were spatially averaged using the LOESS technique \citep{Cleveland1988LocallyWR} implemented by \cite{2013MNRAS.432.1862C}.}
         \label{fig:eps_lam_loess_paper3}
    \end{figure*}

    \begin{figure*}
    \hspace{1cm}
    \section{General properties and spin parameter}
    \centering
       \includegraphics[width=8.5cm]{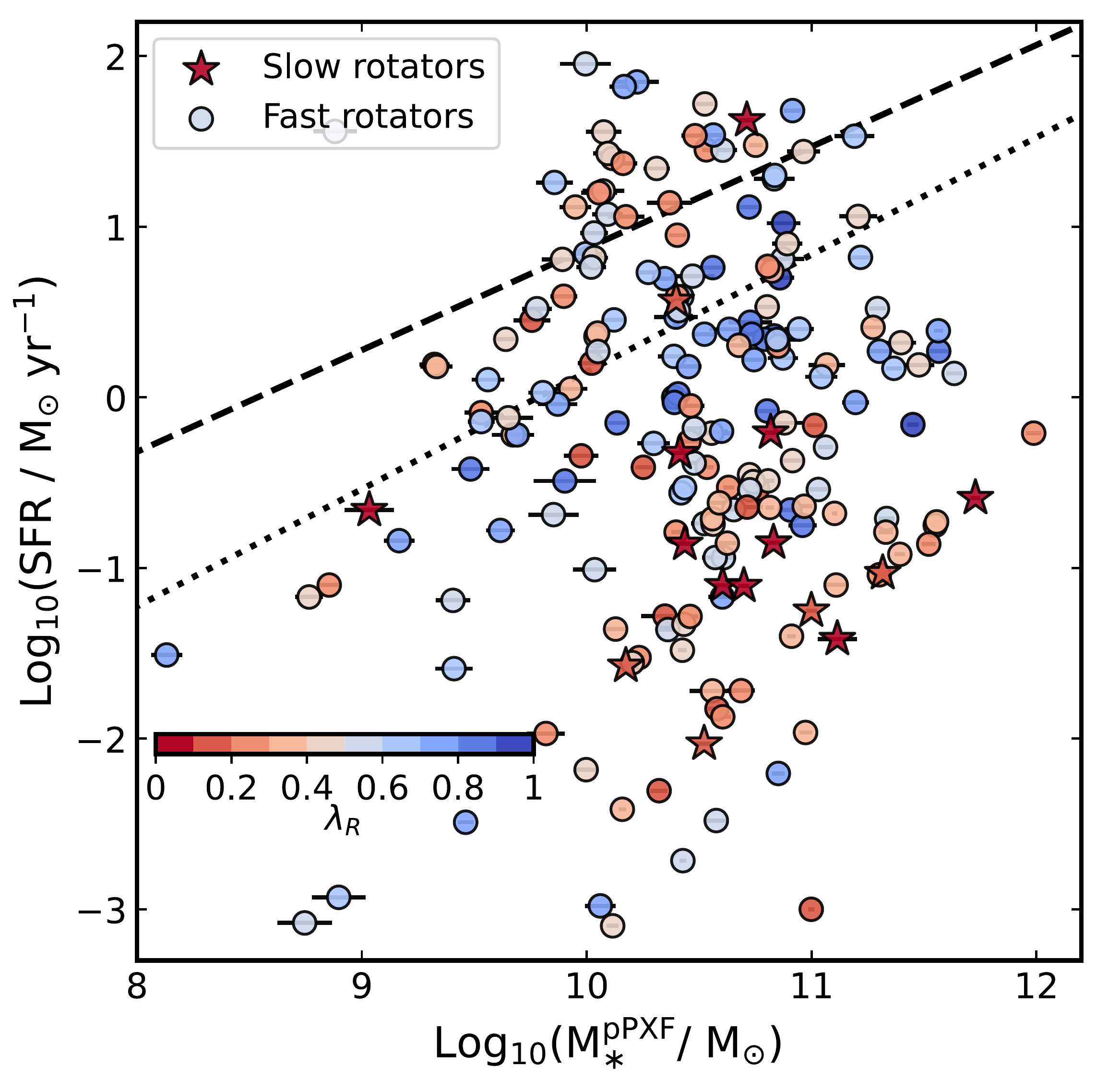}
         \caption{Star formation rate as a function of the stellar mass (derived from the MUSE spectra) for the galaxy sample, coloured coded by the  spin parameter. Stars and circles show slow and fast rotators, respectively. The dotted and dashed black lines indicate the main sequence (MS) for an age of the Universe respectively of 12.4 Gyr (z = 0.1) and 6.8 Gyr (z = 0.8) repectively, given by Eq. (1) of \cite{2017ApJ...838...19W}.}
         \label{fig:mass_sfr_lam_paper3}
    \end{figure*}

\onecolumn

\section{Galaxy's properties}\label{sec:sample_table_paper3}

\begin{table}[hbt!]
\centering
\begin{tabular}{ |p{3cm}||p{11.5cm}|  }
    \hline
    \multicolumn{2}{|c|}{Galaxy sample properties} \\
    \hline
    \hline
    Parameter& Description\\
    \hline
    \hline
    N  &  Galaxy index\\
    survey & Galaxy's survey\\ 
    RA  &  RA J2000 degrees of galaxy centre\\
    DEC  &  Dec J2000 degrees of galaxy centre\\
    Z  &  MUSE spectroscopic redshift\\
    pa & Position angle\\
    vel$\_$disp & Galaxy's velocity dispersion measured within 1 R$_{e}$\\
    lam$\_$obs & Observed spin parameter within the selected aperture\\
    lam$\_$corr  & Corrected spin parameter within the selected aperture \\
    correct$\_$psf & $C$ or $NC$ if the galaxy was o was not seeing corrected, respectively\\
    lam$\_$error &  Error in the spin parameter\\
    aper$\_$selected & Selected aperture for the computation of the spin parameter\\
    kin$\_$class & kinematic classification\\ 
    FR$\_$SR & Fast rotator (FR) or slow rotator (SR)\\
    res$\_$param & Resolution parameter ($r$) given by $r$ = $\sigma_{\rm{PSF}}/ \rm{R^{maj}}$\\
    Re$\_$maj$\_$kpc  &  Size parameter R$^{\rm maj}$ in kpc\\
    lmass$\_$ppxf  &  Galaxy stellar mass from spectral pPXF fitting\\
    lmass$\_$ppxf$\_$err  &  Galaxy stellar mass error from spectral pPXF fitting\\
    lage  &   Galaxy stellar ages from spectral pPXF fitting\\
    lage$\_$err  &  Galaxy stellar ages error from spectral pPXF fitting\\
    metal   &   Galaxy stellar metallicity from spectral pPXF fitting\\
    metal$\_$err   &  Galaxy stellar metallicity error from spectral pPXF fitting\\
    t$\_$90   &  Time taken to assemble 90$\%$ of the stellar mass\\
    eta   &  Environmental global density parameter\\
    N$\_$SFE   & Number of star-forming episodes\\\hline
\end{tabular}
\caption{Galaxy sample properties. The full Table is available at the CDS.}
\label{tab:sample_table} 
\end{table}

\end{appendix}
\end{document}